\documentclass{aa}  

\usepackage{graphicx}
\usepackage{txfonts}
\begin{document}

   \title{The XMM-Newton serendipitous survey\thanks{Based on observations obtained with XMM-Newton, an ESA science mission with instruments and contributions directly funded by ESA Member States and NASA.}}

   \subtitle{VII. The third XMM-Newton serendipitous source catalogue\thanks{http://cdsarc.u-strasbg.fr/viz-bin/VizieR?-meta.foot\&-source=IX/44}}

   \author{S. R. Rosen
          \inst{1} \and
          N. A. Webb
          \inst{2,3}
          \and
          M. G. Watson\inst{1}
          \and
          J. Ballet\inst{4} \and D. Barret\inst{2,3} \and V. Braito\inst{1,6}
          \and F. J. Carrera\inst{5} \and M. T. Ceballos\inst{5} \and
          M. Coriat\inst{2,3} \and R. Della Ceca\inst{6} \and
          G. Denkinson\inst{1} \and P. Esquej\inst{1} \and
          S. A. Farrell\inst{1} \and M. Freyberg\inst{7} \and
          F. Gris\'e\inst{8} \and  P. Guillout\inst{8} \and L. Heil\inst{1}
          \and F. Koliopanos\inst{2,3} \and D. Law-Green\inst{1} \and G. Lamer\inst{10} \and
          D. Lin\inst{2,3,12} \and R. Martino\inst{4} \and L. Michel\inst{8}
          \and C.  Motch\inst{8} \and A. Nebot Gomez-Moran\inst{8} \and C. G. Page\inst{1} \and K. Page\inst{1} \and M. Page\inst{11} \and M.W. Pakull\inst{8} \and  J. Pye\inst{1} \and A. Read\inst{1} \and P. Rodriguez\inst{9} \and M. Sakano\inst{1} \and R. Saxton\inst{9} \and A. Schwope\inst{10} \and A. E. Scott\inst{1} \and R. Sturm\inst{7} \and I. Traulsen\inst{10} \and V. Yershov\inst{11} \and I. Zolotukhin\inst{2,3} }

   \institute{Department of Physics \& Astronomy, University of Leicester
, Leicester, LE1 7RH, UK
            \and  
            Universit\'e de Toulouse; UPS-OMP, IRAP,  Toulouse, France\\
              \email{Natalie.Webb@irap.omp.eu}
         \and
             CNRS, IRAP, 9 av. Colonel Roche, BP 44346, F-31028 Toulouse cedex 4, 
France
\and
Laboratoire AIM, CEA-IRFU/CNRS/Universit\'e Paris Diderot, Service d’Astrophysique, CEA Saclay, 91191 Gif sur Yvette, France
   \and
Instituto de Fisica de Cantabria (CSIC-UC), Avenida de los Castros, 39005 Santander, Spain 
\and
INAF-Osservatorio Astronomico di Brera, via Brera 28, I-20121 Milano, Italy 
\and
Max-Planck-Institut f\"ur extraterrestrische Physik, Giessenbachstr., 85748 Garching, Germany
\and
Observatoire astronomique, Universit\'e de Strasbourg, CNRS, UMR 7550, 11 rue de l'Université, F-67000 Strasbourg, France
\and
XMM SOC, ESAC, Apartado 78, 28691 Villanueva de la Cañada, Madrid, Spain
\and
       Leibniz-Institut f\"ur Astrophysik Potsdam (AIP), An der Sternwarte 16, 14482 Potsdam, Germany 
\and
Mullard Space Science Laboratory, University College London, Holbury St Mary, Dorking, Surrey RH5 6NT, UK       
\and 
Space Science Center, University of New Hampshire, 8 College Road, Durham, NH 03824-2600, U.S.A.        }

   \date{Received ; accepted }

 
  \abstract
   {Thanks to the large collecting area (3 $\times \sim$1500 cm$^2$ at 1.5
     keV) and wide field of view (30\arcmin\ across in full field mode) of the
     X-ray cameras on board the European Space Agency X-ray observatory {\em
       XMM-Newton}, each individual pointing can result in the detection of up
     to several hundred X-ray sources, most of which are newly discovered
     objects.  Since {\em XMM-Newton} has now been in orbit for more than 15 years, hundreds of thousands of sources have been detected.   } 
   {Recently, many improvements in the {\em XMM-Newton} data reduction algorithms have been made. These include enhanced source characterisation and reduced spurious source detections, refined astrometric precision of sources, greater net sensitivity for source detection, and the extraction of spectra and time series for fainter sources, both with better signal-to-noise. Thanks to these enhancements, the quality of the catalogue products has been much improved over earlier catalogues. Furthermore, almost 50\% more observations are in the public domain compared to 2XMMi-DR3, allowing the {\em XMM-Newton Survey Science Centre} to produce a much larger and better quality X-ray source catalogue.}
   {The {\em XMM-Newton Survey Science Centre} has developed a pipeline to reduce the {\em XMM-Newton} data automatically. Using the
     latest version of this pipeline, along with better calibration, a new
     version of the catalogue has been produced, using {\em
       XMM-Newton} X-ray observations made public on or before 2013 December 31. Manual screening of all of the X-ray detections ensures the highest data quality.  This catalogue is known as 3XMM.}
   {In the latest release of the 3XMM catalogue, 3XMM-DR5, there are 565962
     X-ray detections comprising 396910 unique X-ray sources. Spectra and lightcurves are provided for the 133000 brightest sources. For all detections, the positions on the sky, a measure of the
     quality of the detection, and an evaluation of the X-ray variability is
     provided, along with the fluxes and count rates in 7 X-ray energy bands,
     the total 0.2-12~keV band counts, and four hardness ratios. With the aim of identifying the detections, a cross correlation with 228 catalogues of sources detected in all wavebands is also provided for each X-ray detection.}
   {3XMM-DR5 is the largest X-ray source catalogue ever produced. Thanks to the large array of data products associated with each detection and each source, it is an excellent resource for finding new and extreme objects.}


   \maketitle
%

\section{Introduction}
{\em XMM-Newton} \citep{jans01} is the second cornerstone mission from the {\em European
  Space Agency Horizon 2000 programme}. It was launched in December 1999, and 
thanks to the $\sim$1500 cm$^2$ of geometric effective area \citep{turn01} for each of the three X-ray 
telescopes aboard, it has the largest effective area of any
X-ray satellite \citep{long14}. This fact, coupled with the large field of view (FOV) of
30\arcmin, means that a single pointing on average detects 50 to 100
serendipitous X-ray sources \citep{wats09}.

For the past 19 years, the {\em XMM-Newton Survey Science Centre}\footnote{http://xmmssc.irap.omp.eu/} (SSC), a consortium of ten
European Institutes \citep{wats01} has developed much of the {\em XMM-Newton
  Science Analysis Software} (SAS) \citep{sas04} for reducing and analysing {\em
  XMM-Newton} data and created pipelines to perform standardised routine 
processing of the {\em XMM-Newton} science data. The XMM SSC 
has also been responsible for producing catalogues of all of the sources
detected with {\em XMM-Newton}. The catalogues of X-ray sources detected with
the three EPIC \citep{stru01,turn01} cameras that are placed at the focal
point of the three X-ray telescopes have been designated 1XMM and 2XMM
successively \citep{wats09}, with incremental versions of these catalogues
indicated by successive data releases, denoted -DR in association with the
catalogue number. This paper presents the latest version of the XMM catalogue,
3XMM. The original 3XMM catalogue was data release 4 (DR4). The publication of 
this paper coincides with the release of 3XMM-DR5. This version includes one extra 
year of data and increases the number of detections by 7\%, with respect to 
3XMM-DR4. The number of X-ray detections in 3XMM-DR5 is 565962, which translate 
to 396910 unique X-ray sources.   The median flux of these X-ray sources is
$\sim$2.4$\times$ 10$^{-14}$ erg cm$^{-2}$ s$^{-1}$ (0.2-12.0 keV), and the data
taken span 13 years. The catalogue covers 877 square degrees of sky 
($\sim$2.1\% of the sky) if the overlaps in the catalogue are taken into account. 
3XMM-DR5 also includes a number of enhancements 
with respect to the 3XMM-DR4 version, which are described in 
appendix~\ref{ap:knownissues}. The 3XMM-DR5 catalogue is approximately 60\% 
larger than the 2XMMi-DR3 release and five times the current size of 
the {\em Chandra} source
catalogue \citep{evan10}. 3XMM uses significant improvements to the SAS and incorporates developments with the calibration. Enhancements include
better source characterisation, a lower number of spurious source
detections, better astrometric precision, greater net sensitivity and spectra,
and time series for fainter sources, both with better signal-to-noise.  These
improvements are detailed throughout this paper.

A complimentary catalogue of ultra-violet and optical sources detected with the {\em
  XMM-Newton} Optical Monitor \citep[OM][]{maso01} in similar fields to the {\em XMM} catalogue is also produced in the framework of the {\em XMM-Newton} SSC and is called the {\em XMM-Newton Serendipitous Ultraviolet Source Survey} \citep[XMM-SUSS in its
  original form, with the more recent version named XMM-SUSS2,][]{page12}.  3XMM is also complementary to other recent X-ray catalogues, such as the {\em Chandra}
source catalogue mentioned above, and the 1SXPS (Swift-X-ray Telescope (XRT) point
source) catalogue \citep{evan14} of 151\hspace*{0.05cm}524 X-ray point sources
detected with the Swift-XRT over eight years of operation. 1SXPS has a sky
coverage nearly 2.5 times that of 3XMM, but the effective area of the XRT
is less than a tenth of each 
of the telescopes on board {\em XMM-Newton} \citep{long14}. Other earlier catalogues include  all-sky coverage, such as the {\em ROSAT all-sky survey}
\citep[RASS][]{voge99}, but the reduced sensitivity of {\em ROSAT} compared to {\em XMM-Newton} means that the RASS catalogue contains just 20\% of the number of sources in 3XMM-DR4. However, the different X-ray source catalogues in conjunction with 3XMM allow searches for
long-term variability. 

Whilst this paper covers the 3XMM catalogue in general, some of the data validation presented was carried out on the 
3XMM-DR4 version that was made public on 23 July 2013. 3XMM-DR4 contains 531261
X-ray detections that relate to 372728 unique X-ray sources taken from 7427
{\em XMM-Newton} observations. 

The paper is structured as follows. Section~\ref{sec:catobs} contains information concerning the observations used in the 3XMM-DR5 catalogue. Section~\ref{sec:processing} covers the 3XMM data processing and details changes made with respect to previous catalogues \citep[see][]{wats09}, such as the exposure selection, the time-dependent boresight implemented, the suppression of minimum ionising particle (MIP) events, the optimised flare filtering, the improved point spread function (PSF) used for the source detection, new astrometric corrections and the newly derived energy conversion factors (ECFs). We also outline the new source flagging procedure. Section~\ref{sec:SSP} covers the source specific products associated with the catalogue, such as the enhanced extraction methods for spectra and time series and the variability characterisation. Section~\ref{sec:screening} describes the various screening procedures employed to guarantee the quality of the catalogue, and Section~\ref{sec:cat_contruct} outlines the statistical methods used for identifying unique sources in the database. Then, Section~\ref{sec:catxcorr} describes the procedures used to cross-correlate all of the X-ray detections with external catalogues, Section~\ref{sec:knownprobs} discusses the limitations of the catalogue and Section~\ref{sec:CATCHAR} characterises the enhancement of this catalogue with respect to previous versions, with the potential of the catalogue highlighted by several examples of objects that can be found in 3XMM, in Section~\ref{sec:examples}. Finally, information on how to access the catalogue is given in Section~\ref{sec:access}, and future catalogue updates are outlined in Section~\ref{sec:updates}, before concluding with a Summary.

\section{Catalogue observations}\label{sec:catobs}
\begin{figure}
   \centering
   \includegraphics[width=9cm]{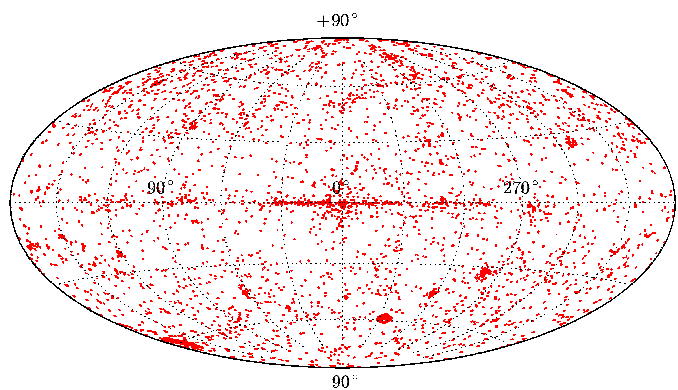}
      \caption{Hammer-Aitoff equal area projection in Galactic coordinates of
the 7781 3XMM-DR5 fields.
              }
         \label{fig:hammer_aitoff}
   \end{figure}

\begin{figure}
   \centering
   \includegraphics[width=9cm]{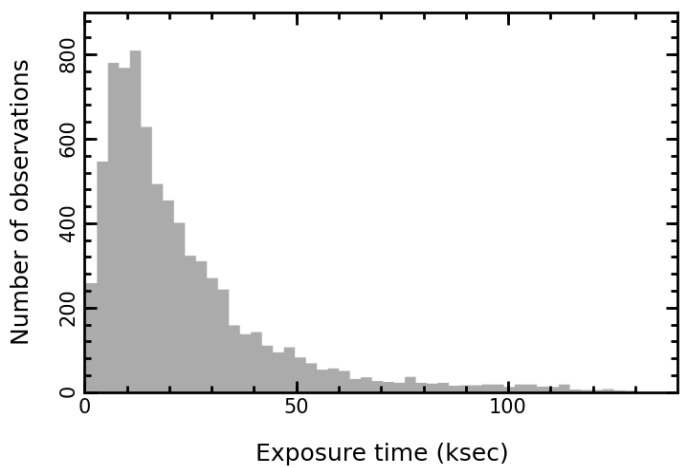}
      \caption{Distribution of total good exposure time (after event filtering) for
the observations included in the 3XMM-DR5 catalogue (for each observation
the maximum time of all three cameras per observation was used).
              }
         \label{fig:MaxExpTime}
   \end{figure}

3XMM-DR5 is comprised of data drawn from 7781 {\em XMM-Newton} EPIC
observations that were publicly available as of 31 December 2013 and that
processed normally. The Hammer-Aitoff equal area projection in Galactic coordinates of
the 3XMM-DR5 fields can be seen in Fig.~\ref{fig:hammer_aitoff}.  The data in 3XMM-DR5
include 440 observations that were publicly available at the time of creating
2XMMi-DR3, but were not included in 2XMMi-DR3 due to the high background or processing
problems. All of those observations containing $>$~1ks clean data ($>$1 ks
of good time interval) were retained for the
catalogue. Fig.~\ref{fig:MaxExpTime} shows the distribution of
total good exposure time (after event filtering) for the observations included
in  the 3XMM-DR5 catalogue and using any of the {\em thick}, {\em medium} or {\em
  thin} filters, but not the {\em open} filter. There are just three observations with one or more cameras configured with the {\em open} filter. The number of the 7781 {\em XMM-Newton} observations included
in the 3XMM-DR5 catalogue for each observing mode and each filter is given in  Table~\ref{tab:7781obs}. Open filter data were processed but not used in the
source detection stage of pipeline processing. The same {\em XMM-Newton} data 
modes were used as in 2XMM \cite{wats09} and are included in appendix \ref{ap:datamodes} of this paper, for convenience.

The only significant difference was
the inclusion of mosaic mode data. Whilst most {\em XMM-Newton} observations
are performed in pointing mode, where the spacecraft is locked on to a fixed
position on the sky for the entire observation, since revolution 1812
(30 October 2009), a specific mosaic observing mode was introduced in which the
satellite pointing direction is stepped across the sky, taking snapshots at
points (sub-pointings) on a user-specified grid. Data from dedicated mosaic
mode or tracking (mosaic-like) observations are recorded into a single
observation data file (ODF) for the observation.  In previous pipeline processing, 
the pipeline products from the small number of mosaic-like observations were generally 
generated, at best, for a single sub-pointing only. This is because the pipeline
filters data such that only events taken during an interval where the attitude
is stable and centred on the nominal observation pointing direction (within a
3\arcmin\ tolerance), are accepted. Data from some, or all, of the other
sub-pointings were thus typically excluded.  During 2012, the {\em XMM-Newton} 
Science Operations Centre (SOC) devised a scheme whereby the parent ODF of a mosaic mode
observation is split into separate ODFs, one for each mosaic sub-pointing. All
relevant data are contained within each sub-pointing ODF and the nominal
pointing direction is computed for the sub-pointing. This approach is applied
to both formal mosaic mode observations and those mosaic-like/tracking
observations executed before revolution 1812. For a mosaic mode observation,
the first 8 digits of its 10-digit observation identifier (OBS\_ID) are common for the parent
observation and its sub-pointings. However, while the last two digits of the
parent observation OBS\_ID almost always end in 01, for the sub-pointings they
form a monotonic sequence, starting at 31. Mosaic mode sub-pointings are thus
immediately recognisable in having OBS\_ID values whose last two digits are
$\ge$ 31.

To the pipeline, mosaic mode (and mosaic-like) observation sub-pointings are
transparent. No special processing is applied. Each sub-pointing is treated as
a distinct observation. Source detection is performed on each sub-pointing
separately and no attempt is made to simultaneously fit common sources
detected in overlapping regions of multiple sub-pointings. While simultaneous
fitting is possible, this aspect had not been sufficiently explored or tested
during the preparations for the 3XMM catalogues.

\begin{table*}
      \caption[]{Characteristics of the 7781 {\em XMM-Newton} observations included
in the 3XMM-DR5 catalogue.}
         \label{tab:7781obs}
   \begin{tabular}{lccccccc}     
\hline\hline       
Camera & \multicolumn{3}{c}{Modes} & \multicolumn{3}{c}{Filters} & Total\\ 
 & Full$^a$ & Window$^b$ & Other$^c$ & Thin & Medium & Thick & \\
\hline                    
pn & 5853 & 495 & - & 3327 & 2633 & 388 & 6348 \\
MOS1 & 6045 & 1306 & 309 & 3296 & 3774 & 590 & 7660 \\
MOS2 & 6100 & 1341 & 248 & 3303 & 3789 & 597 & 7689 \\
\hline                  
\end{tabular}
\\ $^a$ Prime Full Window Extended (PFWE) and Prime Full Window (PFW) modes;
$^b$ pn Prime Large Window (PLW) mode and any of the various MOS
Prime Partial Window (PPW) modes; $^c$ other MOS modes (Fast Uncompressed
(FU), Refresh Frame Store (RFS)).
   \end{table*}

There are 45 observations performed in the dedicated mosaic mode before the
bulk processing cut-off date of 8 December 2012, of which 37 are included in
3XMM-DR5, see appendix~\ref{ap:knownissues}, point 1. None of 
these was available for catalogues prior to 3XMM.  In total, there are 356 processed mosaic sub-pointings
in the 3XMM-DR5 catalogue.

\section{Data processing}\label{sec:processing}
The data used for the 3XMM catalogues have been reprocessed with the latest
version of the SAS and the most up to date calibration available at the time of the processing. 
The majority of the processing for 3XMM-DR5 was conducted during December 2012/January 2013, with 
the exception of 20 observations processed during 2013. The SAS used was
similar to SAS 12.0.1 but included some upgraded tasks required for the
pipeline. The SAS manifest for tasks used in the cat9.0 pipeline and the static set of current 
calibration files (CCFs) that were used for the bulk reprocessing are provided via a dedicated online 
webpage\footnote{http://xmmssc-www.star.le.ac.uk/public/pipeline/doc/04\_cat9.0\_20121220.153800/det\_04\_cat9.0\_20121220.153800}.

There are 31 observations in 2XMMi-DR3 that did not make it in to 3XMM-DR5,
mainly due to software/pipeline errors during processing. Typical examples of
the latter problems are due to revised ODFs (e.g. with no useful
time-correlation information), more sophisticated SAS software that identified
issues hitherto not trapped, or issues with exposure corrections of background
flare light curves and pn time-jumps.

The main data processing steps used to produce the 3XMM data products were
similar to those outlined in \cite{wats09} and described on the SOC
webpages\footnote{http://xmm.esac.esa.int/sas/current/howtousesas.shtml}.  In
brief, these steps were the production of calibrated detector events from the
ODFs; identification of stable background time
intervals; identification of “useful” exposures (taking account of exposure
time, instrument mode, etc.); generation of multi-energy-band X-ray images and
exposure maps from the calibrated events; source detection and
parameterisation; cross-correlation of the source list with a variety of
archival catalogues, image databases and other archival resources; creation of
binned data products; application of automatic and visual screening
procedures to check for any problems in the data products.  The data from this
processing have been made available through the {\em XMM-Newton} Science
Archive\footnote{http://xmm.esac.esa.int/xsa/} (XSA).

\subsection{Exposure selection}\label{sec:expsel}
The only change applied for identifying exposures to be processed by the
pipeline compared to that adopted in pre-cat9.0 processing (\cite{wats09} 
- see their section 4.1), was the exclusion of any exposure taken with the Open
filter. This was done because use of the Open filter leads to increased
contamination from optical light (optical loading). Eight exposures (from five 
observations) taken with the Open filter were excluded from the data
publicly available for the 3XMM-DR5 catalogue.

\subsection{Event list processing}\label{sec:evliproc}
Much of the pipeline processing that converts raw ODF event file data from the
EPIC instruments into cleaned event lists has remained unchanged from the
pre-cat9.0 pipeline and is described in section 4.2 of
\cite{wats09}. However, we describe 3 alterations to the approach used for
2XMM.

\subsubsection{Time-dependent boresight}\label{sec:tdbore}
Analysis by both the {\em XMM-Newton} SSC and the SOC established the presence
of a systematic, cyclic ($\approx$362~day) time-dependent variation in the 
offset of each EPIC (and OM and RGS)
instrument boresight from their nominal pointing positions, for each
observation.  This seasonal dependence is superposed on a long term trend, the
semi-amplitude of the seasonal oscillation being $\approx$1.2\arcsec\ in the
case of the EPIC instruments \citep{Talavera2012rn}. These variations of the
instrument boresights have been characterised by
simple functions in calibration \citep{Talavera2012rn, Talavera2014rn}. The
origin of the variation is uncertain but might arise from heating effects in
the support structures of the instruments and/or spacecraft star-trackers - 
no patterns have been identified in the available housekeeping 
temperature sensor data though these may not sample the relevant parts of the
structure.

During pipeline processing of {\em XMM-Newton} observations for the 3XMM catalogues, 
corrections for this time-dependent boresight movement are applied to 
individual event positions in each instrument, via the SAS task {\it attcalc}, 
based on the observation epochs of the events.

\subsubsection{Suppression of Minimum Ionizing Particle events in EPIC-pn data}\label{sec:mips}
High energy particles can produce electron-hole pairs in the silicon substrate
of the EPIC-pn detector. While onboard processing and standard pn event
processing in the pipeline removes most of these so-called minimum ionizing
particle events \citep{pninstref}, residual effects can arise when MIPs
arrive during the pre-exposure offset-map analysis and can give rise to
features that appear as low-energy noise in the pn detector. Typically, these
features are spatially confined to a clump of a few pixels and appear only in
band 1. However, in pre-cat9.0 pipeline processing, such features were sometimes
detected as sources during source detection and these were not always
recognised and flagged during the manual flagging process outlined in section
7.4 of \cite{wats09}. The SAS task, {\it epreject}
was incorporated into the pipeline processing for 3XMM and in most cases
corrects for these MIP events during processing of pn events.

\subsubsection{Optimised flare filtering}\label{sec:optflare}
In previous pipeline processing (pre-cat9.0 pipelines), the recognition of
background flares and the creation of good time intervals (GTIs) between them
was as described in section 4.3 of \cite{wats09}, where the background
light curves were derived from high energy data and the count rate thresholds
for defining the GTIs were based on (different) constant values for each
instrument. In the processing for 3XMM, two key changes have been made.

Firstly, rather than adopting fixed count rate thresholds in each instrument,
above which data are rejected, an optimisation algorithm has been applied that
maximises the signal-to-noise (S/N) for the detection of point sources.
Secondly, the light curves of the background data used to establish the count
rate threshold for excluding background flares are extracted in an 'in-band'
(0.5-7.5 keV) energy range. This was done so that the process described below
resulted in maximum sensitivity to the detection of objects in the energy
range of scientific interest.

The overall process for creating the background flare GTIs for each exposure 
within each observation involved the following steps: 
\begin{enumerate}
\item For each exposure, a high 
energy light curve (from 7 to 15 keV for pn, $>$ 14 keV for MOS) is created,
as previously, and initial background flare
GTIs are derived using the optimised approach employed in the SAS task,
{\it bkgoptrate} (see below). 
\item Following the identification of bad pixels, event cleaning and 
event merging from the different CCDs, an in-band image is then created, using
the initial GTIs to excise background flares. 
\item The SAS task, {\it eboxdetect} then  
runs on the in-band image to detect sources with a likelihood $> 15$ - this is 
already very conservative as only very bright (likelihood $\gg 100$), variable
sources are able to introduce any significant source variability component 
into the total count rate of the detector (accumulated from most of the
field). 
\item An in-band light curve is subsequently generated, excluding events from 
circular regions of radius 60\arcsec\ for sources with count rates $\le$0.35
counts/s or 100\arcsec\ for sources with count rates $>$0.35counts/s, centred on 
the detected sources.
\item The SAS task, {\it bkgoptrate}, is then applied to the light curve to find the 
optimum background rate cut threshold and this is subsequently used to define the 
final background flare GTIs.
\end{enumerate}

The optimisation algorithm adopted, broadly follows that used for the
processing of ROSAT Wide Field Camera data for the ROSAT 2RE catalogue 
\citep{Pye1995}. The process seeks to determine the background count rate 
threshold at which the remaining data below the threshold yields a S/N ratio, 
$S = \frac {C_s} {\sqrt{C_b}}$, for a (constant) source that is a maximum. 
Here  $C_s$ is the number of source counts and $C_b$ is the number of 
background counts. Since we are interested, here, in finding the background 
rate cut that yields the maximum S/N and are not concerned about 
the absolute value of that S/N, then for background light curves with bins 
of constant width, as created by the pipeline processing, $S$ can be 
expressed as 

\begin{equation}\label{eq:bgs2n}
S = \frac {N}{\sqrt{\sum{r_i}}  }
\end{equation}

where $N$ is the number of bins with background count rates below the 
threshold, $r_T$, and $r_i$ is the count rate in time bin $i$: the summation 
is over the time bins with a count rate $< r_T$. Time bins are of 10s width
for pn and 26s for MOS. The process sorts the time 
bins in order of decreasing count rate. Starting from the highest count rate 
bin, bins are sequentially removed, computing equation \ref{eq:bgs2n} at each 
step. With the count rate of the bin removed at each step representing a 
trial background count rate cut threshold, this process yields a curve of S/N 
vs. background count rate cut threshold. The background cut corresponding 
to the peak of the S/N curve is thus the optimum cut threshold. 

\begin{figure}
   \centering
   \includegraphics[width=9cm]{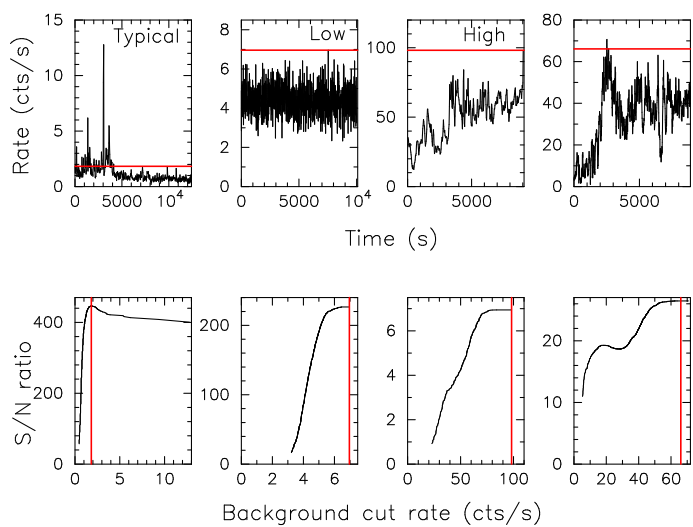}
      \caption{Flare background light curves (top row) and their corresponding S/N vs.
        background cut threshold plots (bottom row). The leftmost panels are
        for a typical observation with notable background flaring. The second pair of
        vertically aligned panels shows an example where the background has a
        persistently low level, while the third pair of panels reflects an
        example where the background is persistently high. The rightmost
        panels show an example of a variable background which gives rise to a
        double-peaked S/N vs. background-rate-cut curve. The vertical red lines in the
        lower panels indicate the optimum background-cut-threshold (i.e. the
        peak of the curve) derived for the light curves in the top panels. In
        the upper panels the applied optimum cut-rate is also shown in red as 
        horizontal lines. 
              }
         \label{fig:example_bkglcs}
   \end{figure}

In figure~\ref{fig:example_bkglcs} we show four examples of in-band background
time series in the top row, accompanied by the respective S/N vs. background-cut-threshold
plots in the bottom row. The first panel in each row represents a typical
observation (MOS1) with some significant background flaring activity. The 
optimum cut level
of 1.83 cts/s leads to the creation of GTIs that exclude portions of the
observation where the background exceeds the cut threshold. The second panels 
are for a pn observation with a stable, low background level. The optimum
cut in the background includes all the data and thus generates a GTI spanning
the entire observation. This is also true for the third panels which show a
MOS1 case where the background is persistently high (above the level where the
whole observation would have been rejected in pre-cat9.0 pipeline
processing). The fourth panels are for an example of a variable
background which gives rise to a double peaked S/N v
background-rate-cut curve. Here, raising the threshold from
$\sim$18 cts/s to $\sim$28 cts/s simply involves a steeply rising background
rate early in the observation, causing a dip in the S/N verses background-rate-cut
curve. However, as the rate cut threshold is increased above 30
cts/s, although the count rate is higher, a lot more exposure time is
available, so the S/N curve rises again and the optimum cut includes almost
all the data. It should be emphasised that the fixed cut thresholds used for
MOS and pn in previous XMM processings can not be directly compared to the
optimised ones used here because of the change in energy band being used to
construct the background light curve. It is, however, worth noting that the
fixed cuts used previously often result in very similar GTIs to those
generated by the optimisation process described above. This is 
because the previous fixed instrument thresholds were based on analyses that 
sought to find a representative level for the majority of {\em XMM-Newton} 
observations. 

We discuss some of the gains of using this optimisation approach 
in section~\ref{sec:CATCHARoptflare} and some known issues in 
section~\ref{sec:knownprobs}.

\subsection{Source detection using the empirical Point Spread Function (PSF)  fitting}\label{sec:PSF}
The bulk reprocessing for 3XMM took advantage of new developments related 
to the EPIC PSFs. The source detection stage in previous pipelines
(\cite{wats09} - see their section 4.4.3) made use of the so-called 'default' (or
medium accuracy) PSF functions determined by ray tracing of the {\em XMM-Newton}
mirror systems. However, these default PSF functions recognised no azimuthal 
dependence in the core of the source profile, did not adequately describe the
prominent spoke structures seen in source images (arising from the mirror
support structures) and were created identically for each EPIC camera.

To address the limitations of the default EPIC PSFs, a set of empirical PSFs
were constructed, separately for each instrument, by careful stacking of
observed XMM source images over a grid of energy and off-axis angles from the
instrument boresights. The cores and spoke patterns of the PSFs were then
modelled independently so that implementation within the {\em XMM-Newton} SAS
calibration software then enables PSFs to be reconstructed that take the off-axis and azimuthal locations of a source into account, as well as the
energy band.  The details of the issues associated with the default PSF and
the construction and validation of the empirical PSF are presented in
\cite{Read2011}.

The use of the empirical PSF has several ramifications in source detection.
Firstly, the better representation of structures in the real PSF results in
more accurate source parameterisation. Secondly, it helps reduce the number of
spurious detections found in the wings of bright sources. This is because the
previous medium accuracy PSFs did not adequately model the core and spoke
features, leaving residuals during fitting that were prone to being detected
as spurious sources. With the empirical PSFs, fewer such spurious detections
are found, especially in the wings of bright objects positioned at larger ($>$
6\arcmin) off-axis angles. Thirdly, as a result of the work on the PSFs, the 
astrometric accuracy of {\em XMM-Newton} source positions has been
significantly improved \citep[see][]{Read2011}.

\subsubsection{Other corrections related to the PSF} \label{sec:otherPSF}
During the late stages of testing of the pipeline used for the bulk
reprocessing that fed into the 3XMM-DR4 catalogue, an analysis of 
{\em XMM-Newton} X-ray source positions relative to the high-accuracy 
($\le 0.1$\arcsec) reference positions of Sloan Digital Sky Survey (SDSS, DR9) quasars identified a small but
significant, off-axis-angle-dependent position shift, predominantly along the
radial vector from the instrument boresight to the source.  
The effect, where the real source position is closer to the instrument
boresight than that inferred from
the fitted PSF centroid, has a negligible displacement on axis and grows
to $\sim$0.65\arcsec\ at off-axis angles of 15\arcmin. This radial shift is due
to the displacement between the true position of a source and the defined
centroid (as determined by a 3-dimensional, circular, Gaussian fit to the
model PSF profile) of the empirical PSF, which grows as the PSF becomes
increasingly distorted at high off-axis angles. It should be noted that
identifying and measuring this effect has only been possible because of the
corrections for other effects (see section~\ref{sec:PSF} and below) that 
masked it, and because of the large number of sources available that provide 
sufficient statistics. In due course a correction for this effect will be 
applied directly to event positions, on a per-instrument basis, via 
the {\em XMM-Newton} calibration system,
but for the 3XMM-DR4 catalogue, to avoid delays in its production, a
solution was implemented within the {\it catcorr} SAS task. A correction,
computed via a third-order polynomial function, is applied to the initial
PSF-fitted coordinates of each source output by {\it emldetect}, i.e. prior to
the field rectification step, based on the off-axis angle of the source as
measured from the spacecraft boresight. This correction is embedded in the 
{\it RA} and {\it DEC} columns, which also include any rectification
corrections (section~\ref{sec:rectification}). The correction is computed and 
applied in the same way for both the 3XMM-DR4 and 3XMM-DR5 catalogues.

A second PSF-related problem that affected 2XMMi-DR3 positions was uncovered
during early testing of the empirical PSF (see Read et al. 2011). This arose
from a 0.5 pixel error (in both the x and y directions) in the definition of
the pixel coordinate system of the medium-accuracy PSF map - as pixels in the 
PSF map are defined to be 1\arcsec\ x 1\arcsec, the error is equivalent to 
0.5\arcsec\ in each direction. When transferred to
the image frame during PSF fitting in {\it emldetect}, this error in the PSF map
coordinate system manifested itself as an offset of up to 0.7\arcsec\ in
the RA/DEC of a source position, varying with azimuthal position within the
field. The introduction of the empirical PSF removes this error.

\subsection{Astrometric rectification}\label{sec:rectification}
\subsubsection{Frame correction}\label{sec:rectfrcorr}
Celestial coordinates of sources emerging from the PSF fitting step of 
pipeline processing of a given observation include a generally small 
systematic error arising from offsets in the spacecraft boresight position 
from the nominal pointing direction for the observation. 
The uncertainty is due to imprecisions in the attitude solution derived from data from the 
spacecraft's star-trackers and may result in frame shifts that are 
typically $\sim$1\arcsec\ (but can be as much as 10\arcsec\ in a few cases) 
in the RA and DEC directions and a rotation of the field about the boresight of 
the order of 0.1 degrees. To correct for (i.e. 
rectify) these shifts, an attempt is made to cross-correlate sources in the 
{\em XMM-Newton} field of view with objects from an astrometric reference catalogue. 
X-ray sources with counterparts in the reference catalogue are used to 
derive the frame shifts and rotation that minimise the displacements  
between them. In all previous pipeline processing (and catalogues derived 
from them) these frame corrections were estimated using the SAS task, 
{\it eposcorr}, which used a single reference catalogue, USNO-B1.0, and the 
SAS task, {\it evalcorr}, to determine the success and reliability of the 
outcome (\cite{wats09} - see their section 4.5). 

The processing system used to create the data for the 3XMM catalogues makes
use of some important 
improvements to this field rectification procedure, which are embedded in the 
new SAS tasks, {\it catcorr} that replaces {\it eposcorr} and {\it evalcorr}.
Firstly, the new approach incorporates an iterative fitting function
\citep{nm1965} to find the optimum frame-shift corrections: previously the 
optimum shift was obtained from a grid-search procedure. Secondly, the
cross-match between {\em XMM-Newton} and reference catalogue source 
positions is carried out using three reference catalogues: (1) USNO-B1.0
\citep{Monet2003}, (2) 2MASS \citep{2mass2006} and, where sky
coverage permits, (3) the Sloan
Digital Sky Survey (DR9)  \citep{Abazajian2009}. The 
analysis is conducted using each catalogue separately. When
there is an acceptable fit from at least one 
catalogue, the RA and DEC frame shifts and the rotation derived from the 
'best' case are used to correct the source positions. A fit is considered
acceptable if there are at least 10 X-ray/counterpart pairs, the maximum
offset between a pair (X-ray source, $i$ and counterpart, $j$) 
is $< 10$\arcsec\ and the goodness of fit statistic

\begin{equation}\label{eq:L}
L = \sum\limits_{i=1}^{n_x} \sum\limits_{j=1}^{n_o}  max(0.0,p_{ij} - q_{ij})
  \ge 5 
\end{equation}

where $p_{ij} = e^{-\frac{1}{2} (r_{ij}/\sigma_{ij})^2}$ and $q_{ij} = n_o
(r_{ij}/r_f)^2$. Here, $p_{ij}$ is the probability of finding the 
counterpart at a distance $ > r_{ij}$ from the X-ray source position given the combined (in
quadrature) positional uncertainty, $\sigma_{ij}$, while $q_{ij}$ is the
probability that the counterpart is a random field object within $r_{ij}$. An
estimate of the local surface density of field objects from the reference
catalogue is made by counting the number, $n_o$, of such objects within a circular
region of radius $r_f$ (set to 1\arcmin) around each XMM source. $n_x$ is the
number of X-ray sources in the XMM field. The $L$ statistic, which represents a
heuristic approach to the problem of identifying likely matching counterparts, is 
computed over the set of matching pairs and is a measure of the
dominance of the closeness of the counterparts over the
probability of random matches. The shifts in RA and DEC and the rotation are
adjusted within the fitting process to maximise $L$. Extensive trials found 
that if $L \ge 5$, the result is generally reliable. Where more than one 
reference catalogue gives an acceptable solution, the one with the largest $L$ 
value is adopted. 

In the 3XMM catalogues, the corrected coordinates are placed in the RA and 
DEC columns; the original uncorrected coordinates are reported via the RA\_UNC 
and DEC\_UNC columns. A catalogue identifier for the catalogue yielding the 
'best' result is provided in the REFCAT column. If the best fit has parameter 
values (e.g. the number of matches used) that fall below the specific
constraints mentioned above, the original, uncorrected positions are 
retained (written to both the RA and DEC and RA\_UNC and DEC\_UNC columns) 
and the REFCAT identifier takes a negative value. Further details may be 
found in the documentation for the {\it catcorr} task. This new rectification 
algorithm is successful for about 83\% of observations, which contain 89\% of 
detections, reflecting a significant improvement compared to the previous 
approach where $\sim$ 65\% of fields could be corrected. The main gain comes 
from the use of the 2MASS catalogue which is particularly beneficial in 
obtaining rectification solutions in the galactic plane - it should be pointed
out that similar gains would be obtained with {\it eposcorr} if used with the
expanded set of reference catalogues. It should be noted that the extracted lists 
of objects from each of the three reference catalogues that lie within the full 
EPIC field of view for a given observation, are provided to users of {\em XMM-Newton} 
data products via the file-type=REFCAT product file, which is used by the 
task, {\it catcorr}.

\subsubsection{Systematic position errors}\label{sec:rectsyserr}

As discussed in section 9.5 of \cite{wats09}, for the 2XMM catalogue (and relevant
to subsequent incremental catalogues in the 2XMM series), the angular deviations 
of SDSS (DR5) quasars \citep{Schneider2007} from their {\em XMM-Newton} X-ray 
counterparts, normalised by the combined position errors, could not be modelled
by the expected Rayleigh distribution unless an additional systematic
uncertainty (SYSERR parameter in 2XMM) was added to the statistical 
position error (RADEC\_ERR parameter in 2XMM and see Appendix~\ref{ap:defs}) derived during the PSF fitting
process. \cite{wats09} showed that this systematic was not consistent with
the uncertainty arising from the rectification procedure used for the 2XMM 
processing and ultimately adopted an empirically-determined systematic error 
value that produced the best match between the distribution of XMM-quasar 
offsets and the expected Rayleigh curve.

As part of the upgrade applied to the rectification process for the bulk
reprocessing used for the 3XMM catalogues, the uncertainty arising from this
step has been computed, in particular, taking into account the error component
arising from the rotational offset. Errors (1$\sigma$) in each component,
i.e., on the RA offset, $\Delta\alpha_c$, on the DEC offset ($\Delta\delta_c$)
and on the rotational angle offset ($\Delta\phi_c$), have been combined in
quadrature to give an estimate of the total positional uncertainty,
$\Delta{r}$, arising from the rectification process as

\begin{equation}
\Delta{r} = [ (\Delta\alpha_c)^2 + (\Delta\delta_c)^2 +
  (\theta_{c}.\Delta\phi_c)^2 ]^{\frac{1}{2}}
\end{equation}

where $\theta_c$ is the radial off-axis angle, measured in the same units 
as $\Delta\alpha_c$ and $\Delta\delta_c$ and $\Delta\phi_c$ is in radians.

Inclusion of this rectification error (column {\it SYSERRCC} in the 3XMM 
catalogues, see Appendix~\ref{ap:defs}), in quadrature 
with the statistical error, leads to a generally good agreement between 
the XMM-quasar offset distribution and the expected Rayleigh distribution 
compared to the previous approach and indicates that the empirically-derived 
systematic used in pre-3XMM catalogues is no longer needed. This is 
discussed further in section~\ref{sec:CATCHARastrometry}.

\subsection{Energy Conversion Factors (ECFs)}\label{sec:ecfs}
A number of improvements in the calibration of the MOS and pn instruments have
occurred since the previous, 2XMMi-DR3, catalogue was produced, which lead to
slight changes in the energy conversion factors (ECFs) that are used for
converting count rates in the EPIC energy bands to fluxes (see \cite{wats09}
section 4.6 and see Appendix~\ref{ap:defs}). Of note is the fact that MOS redistribution matrices were
provided for 13 epochs at the time of processing for 3XMM and for three areas
of the detector that reflect the so-called 'patch', 'wings-of-patch' and
'off-patch' locations \citep{sembay2011}.

For the 3XMM catalogues a simple approach has been adopted. ECFs were 
computed following the prescription of \cite{mateos09}, for energy bands 1 to 5
(0.2-0.5~keV, 0.5-1.0~keV, 1.0-2.0~keV, 2.0-4.5~keV and 4.5-12.0~keV respectively) and 
band 9 (0.5-4.5~keV), for full-frame mode, for each EPIC camera, for each of the Thin, Medium and
Thick filters. A power-law spectral model with a photon index,
$\Gamma = 1.7$ and a cold absorbing column density of $N_H = 3\times
10^{20}$~cm$^{-2}$ was assumed. As such, users are reminded that the ECFs, and
hence the fluxes provided in the 3XMM catalogues, may not accurately reflect
those for specific sources whose spectra differ appreciably from this power-law 
model - see section 4.6 of \cite{wats09}.

For pn, the ECFs are calculated at the on-axis position. The pn
response is sufficiently stable that no temporal resolution is needed.
For MOS, to retain a direct connection between the ECFs and publicly available
response files, the ECFs used are taken at epoch 13 and are for the
'off-patch' location. The latter choice was made because the large majority of
detections in an {\em XMM-Newton} field lie outside the 'patch' and 'wings-of-patch'
regions, which only relate to a region of radius $\le$ 40\arcsec, near the
centre of the field. The use of a single epoch (epoch 13) was made to retain
simplicity in the processing and because the response of the MOS cameras
exhibits a step function change (due to a gain change) between epochs 5 and 6, with different but
broadly constant values either side of the step. None of the 13 calibration
epochs represent the average response and thus no response file exists to
which average ECFs can be directly related. The step-function change in the
responses for MOS is most marked in band 1 (0.2-0.5 keV) for the 'patch'
location, where the maximum range in ECFs either side of the step amounts to
20\%. Outside the 'patch' region, and for all other energy bands, the range of
the ECF values with epoch is $\le$ 5\% and is $\le$ 2.5\% for the 'off-patch'
region. Epoch 13 was chosen, somewhat arbitrarily, as being typical of epochs
in the longer post-step time interval.

The ECFs, in units of $10^{11}$~cts~cm$^2$~erg$^{-1}$, adopted for the bulk 
reprocessing of data used for 3XMM, are provided in Table~\ref{tab:ecfs}, for 
each camera, energy band and filter. The camera rate, {\it ca\_RATE}, and 
flux, {\it ca\_FLUX}, are related via {\it ca\_FLUX} = ({\it{ca\_RATE}}/ECF) 
(where {\it ca} is PN, M1 or M2) 

\begin{table}[ht]
\begin{minipage}[ht]{\columnwidth}
\normalsize
\caption{Energy conversion factors (in units of $10^{11}$~cts~cm$^2$~erg$^{-1}$)
used to convert count rates to fluxes for each instrument, filter and energy band}
\label{tab:ecfs}
\small
\centering
\renewcommand{\footnoterule}{} 
\tabcolsep 0mm
\begin{tabular}{
l @{\extracolsep{2mm}} c @{\extracolsep{5mm}} 
c @{\extracolsep{6mm}} c @{\extracolsep{6mm}} c }
\hline \hline
       &      & \multicolumn{3}{c}{Filters}       \\
Camera & Band & \multicolumn{1}{c}{Thin} &
\multicolumn{1}{c}{Medium} & \multicolumn{1}{c}{Thick} \\
\hline
pn	&  1 & 9.52 & 8.37 & 5.11 \\
	&  2 & 8.12  & 7.87 & 6.05 \\
	&  3 & 5.87  & 5.77 & 4.99 \\
	&  4 & 1.95 & 1.93  & 1.83 \\
	&  5 & 0.58 & 0.58 & 0.57 \\
	&  9 & 4.56 & 4.46 & 3.76 \\
MOS1	&  1 & 1.73 & 1.53 & 1.00 \\
	&  2 & 1.75 & 1.70 & 1.38 \\
	&  3 & 2.04 & 2.01 & 1.79 \\
	&  4 & 0.74 & 0.73 & 0.70 \\
	&  5 & 0.15  & 0.15 & 0.14  \\
	&  9 & 1.38 & 1.36 & 1.20 \\
MOS2	&  1 & 1.73 & 1.52 & 0.99 \\
	&  2 & 1.76 & 1.71 & 1.39 \\
	&  3 & 2.04 & 2.01 & 1.79 \\
	&  4 & 0.74 & 0.73 & 0.70  \\
	&  5 & 0.15 & 0.15 & 0.15 \\
	&  9 & 1.39 & 1.36 & 1.21 \\

\hline
\end{tabular}
\end{minipage}
\normalsize
\end{table}

\subsection{Updated flagging procedures}\label{sec:flagging}
\label{sec:qflags}
A significant issue in terms of spurious detections in {\em XMM-Newton} data 
arises
from detections associated with out-of-time (OoT) events. For sources that do
not suffer significantly from pile-up, the background map used by emldetect
includes a component that models the OoT features. However, for sources where
pile up is significant, the OoT modelling is inadequate. This can give rise to
spurious sources being detected along OoT features. For the more piled up
objects, the numbers of spurious detections along OoT features can become
large (tens to hundreds).

Another feature arising from bright sources that affects the MOS instruments
is scattered X-rays from the Reflection Grating Arrays (RGA). These manifest
themselves as linear features in MOS images passing through the bright object,
rather similar in appearance to OoT features. These features are not modelled
at all in the background map.

In previous catalogues, spurious detections associated with OoT and RGA
features have simply been masked during manual screening. In the cat9.0
pipeline, for the first time, an attempt has been made to identify the presence of
OoT and RGA features from piled up sources and to flag detections that are
associated with them.

The SAS task, {\it eootepileupmask}, is used for this purpose. This task uses 
simple instrument (and mode) -dependent pre-defined thresholds to test 
pixels in an image for pile-up. Where it detects pixels that exceed the 
threshold, the column containing that pixel is flagged in a mask map for the 
instrument. The task attempts to identify and mask columns and rows 
associated with such pixels in OoT and RGA features.

Once the pile up masks are generated, the SAS task, {\it dpssflag} is used to 
set flag 10 of the {\it PN\_FLAG, M1\_FLAG, M2\_FLAG, EP\_FLAG} columns in the
catalogues for any detection whose centre lies on any masked column or row.

\section{Source-specific product generation}\label{sec:SSP}
\subsection{Optimised spectral and time series extraction}\label{sec:SSPoptext}

The pipeline processing automatically extracts spectra and time series
(source-specific products, SSPs), from suitable exposures, for detections that
meet certain brightness criteria.

In pre-cat9.0 pipelines, extractions were attempted for any source which had 
at least 500 EPIC counts. In such cases, source data
were extracted from a circular aperture of fixed radius (28\arcsec),
centred on the detection position, while background data were accumulated from
a co-centred annular region with inner and outer radii of 60\arcsec\ and 
180\arcsec, respectively. Other sources that lay within or overlapped the
background region were masked during the processing. In most cases this
process worked well. However, in some cases, especially when extracting SSPs
from sources within the small central window of MOS Small-Window mode
observations, the background region could comprise very little usable
background, with the bulk of the region lying in the gap between the central
CCD and the peripheral ones. This resulted in very small (or even zero) areas
for background rate scaling during background subtraction, often leading to
incorrect background subtraction during the analysis of spectra in XSPEC 
\citep{arnaud96}.

For the bulk reprocessing leading to the 3XMM catalogues, two new approaches 
have been adopted and implemented in the cat9.0 pipeline.

\begin{enumerate}
\item{The extraction of data for the source takes place from an aperture
whose radius is automatically adjusted to maximise the signal-to-noise 
(S/N) of the source data. This is achieved by a curve-of-growth analysis, 
performed by the SAS task, {\it eregionanalyse}. This is especially useful for 
fainter sources where the relative importance of the background level is higher.}

\item{To address the problem of locating an adequately filled background
region for each source, the centre of a circular background aperture of
radius, $r_b=168$\arcsec\ (comparable area to the previously used annulus)
is stepped around the source along a circle centred on the source
position. Up to 40 uniformly spaced azimuthal trials are tested along each
circle. A suitable background region is found if, after masking out other
contaminating sources that overlap the background circle and allowing for 
empty regions, a filling factor of at least 70\% usable area remains. 
If none of the  background region trials along a given
circle yields sufficient residual background area, the background region 
is moved out to a circle of larger radius from the object and the azimuthal 
trials are repeated. The smallest trial circle has a radius, $r_c$, of 
$r_c=r_b + 60$\arcsec\ so that the inner edge of the background region is at 
least 60\arcsec\ from the source centre - for the case of MOS Small-Window 
mode, the smallest test circle for a source in the central CCD is set to a 
radius that already lies on the peripheral CCDs. Other than for the MOS
Small-Window cases, a further constraint is that, ideally, the background
region should lie on the same instrument CCD as the source. If no solution 
is found with at least a 70\% filling factor, the background trial with the 
largest filling factor is adopted.

For the vast majority of detections where SSP extraction is attempted,
this process obtains a solution in the first radial step and a strong
bias to early azimuthal steps, i.e. in most cases an acceptable solution
is found very rapidly. For detections in the MOS instruments, about 1.7\%
lie in the central window in Small-Window mode and have a background
region located on the peripheral CCDs. Importantly, in contrast to earlier
pipelines, this process always yields a usable background spectrum for
objects in the central window of MOS Small-Window mode observations. 

This approach to locating the background region was adopted primarily to
provide a single algorithm that works for all sources, including those 
located in the MOS small window, when used. However, a drawback
relative to the use of the original annular background region arises where sources 
are positioned on a notably ramped or other spatially variable background
(e.g. in the wings of cluster emission), where the background that is 
subtracted can vary, depending on which side of the source the background 
region is located. 
}
\end{enumerate}

In addition, the cat9.0 pipeline permits extraction of SSPs for fainter
sources. Extraction is considered for any detection with at least 100 EPIC
source counts (EP\_8\_CTS). Where this condition is met, a spectrum from the
source aperture (i.e. source plus background) is extracted. If the number of
counts from spectrum channels not flagged as 'bad' (in the sense adopted by 
XSPEC) is > 100, a spectrum and time series are extracted for the exposure. 
The initial filter on EPIC counts is used to limit the processing time as, 
for dense fields, the above background location process can be slow. 

\subsection{Attitude GTI filtering} \label{sec:SSPattfilt}

Occasionally, the spacecraft can be settling on to, or begin moving away from, the 
intended pointing direction within the nominal observing window of a pointed 
{\em XMM-Newton} observation, resulting in notable attitude drift at the start or end of an 
exposure. Image data are extracted from events only within 'Good Time Intervals' (GTIs) 
when the pointing direction is within 3\arcmin\ of the nominal pointing position for the 
observation. However, in pre-cat9.0 pipelines, spectra and time series have been extracted 
without applying such attitude GTI filtering. Occasionally, this resulted in a source 
location being outside or at the edge of the field of view when some events were being 
collected, leading to incorrect transitions in the time series. In some cases, 
these transitions gave rise to the erroneous detection of variability in 
subsequent time series processing. In the cat9.0 pipeline, attitude GTI
filtering is applied during the extraction of spectra and time series. 

\subsection{Variability characterisation}\label{sec:varchar}
As with pre-cat9.0 pipeline processing, the pipeline processing for the 
3XMM catalogues subjects each extracted exposure-level source time series to 
a test for variability. This test is a simple $\chi^2$ analysis for the null 
hypothesis of a constant source count rate (\cite{wats09} - see their section 5.2). 
Sources with a probability $\le 10^{-5}$ of being constant 
have been flagged as being variable in previous {\em XMM-Newton} X-ray 
source catalogues and this same approach is adopted for 3XMM. 

In addition, for 3XMM, we have attempted to characterise the scale of the
variability through the fractional variability amplitude, $F_{var}$ (provided
via the {\it PN\_FVAR, M1\_FVAR} and {\it M2\_FVAR} columns and associated
{\it FVARRERR} columns), which is simply the square
root of the excess variance, after normalisation by the mean count rate, 
$\langle{R}\rangle$, i.e.  

\begin{equation}\label{eq:fvar}
F_{var} = \sqrt{ \frac {(S^2 - \langle{\sigma_{err}}^2 \rangle) } {{\langle{R}\rangle}^2 } }
\end{equation}

(e.g. \cite{Edelson2002, Nandra1997} and references therein), where
$S^2$ is the observed variance of the time series with $N$ bins,  i.e. 


\begin{equation}\label{eq:s}
S^2 = {\frac {1}{N-1}} \sum\limits_{i=1}^N (R_i - \langle{R}\rangle)^2 \nonumber
\end{equation}

in which $R_i$ is the count rate in time bin $i$. For the calculation of the
excess variance, $(S^2 - \langle{\sigma_{err}}^2
\rangle)$, which measures the level of observed variance above that expected
from pure data measurement noise, the noise component, 
$\langle{\sigma_{err}}^2 \rangle$, is computed as the mean of the squares of
the individual statistical errors, $\sigma^2_i$, on the count rates of each 
bin, $i$, in the time series.

The uncertainty, $\Delta(F_{var})$, on $F_{var}$, is calculated following
equation B2 in appendix B of \cite{Vaughan2003}, i.e. 

\begin{equation}\label{eq:fvare}
\Delta(F_{var}) = \Bigg[ \Bigg({\sqrt{\frac{1}{2N}} } { \frac{
      \langle{\sigma_{err}}^2 \rangle }{ {\langle{R}\rangle}^2 F_{var} }  }
  \Bigg)^2       +  \Bigg({\sqrt{\frac{\langle{\sigma_{err}}^2 \rangle}{N}} } { \frac{1}{ {\langle{R}\rangle}}  }
  \Bigg)^2      \Bigg]^\frac{1}{2}  \nonumber
\end{equation}

This takes account of the statistical errors on the time bins but not scatter 
intrinsic to the underlying variability process.

\section{Screening} \label{sec:screening}
As for previous {\em XMM-Newton} X-ray source catalogues (\cite{wats09} - see
section 7), every {\em XMM-Newton} observation in the 3XMM catalogues has 
been visually inspected with the purpose of
identifying problematic areas where source detection or source
characterisation are potentially suspect.  The manual screening process
generates mask files that define the problematic regions. These may be
confined regions around individual suspect detections or larger areas
enclosing multiple affected detections, up to the full area of the field where
serious problems exist. Detections in such regions are subsequently assigned a
manual flag (flag 11) in the flag columns ({\it PN\_FLAG}, {\it M1\_FLAG},
{\it M2\_FLAG}, {\it EP\_FLAG}). It should be noted that a detection with 
flag 11 set to (T)rue does not necessarily indicate that the detection is 
considered to be spurious.

One significant change to the screening approach adopted for 3XMM relates
to the flagging of bright sources and detections within a halo of suspect
detections around the bright source. Previously, all detections in the halo
region, including the primary detection of the bright source itself (where
discernible), had flag 11 set to True (manual flag) but the primary detection
of the bright object itself, also had flag 12 set. The meaning of flag 12
there was to signify that the bright object detection was not considered
suspect. The use of flag 12 in this 'negative' context, compared to the other
flags, was considered to be potentially confusing. For this reason, for the 
3XMM catalogues, we have dropped the use of flag 12 and simply ensured that, 
where the bright object detection is clearly identified, it is un-flagged
(i.e. neither flag 11 or 12 are set). Effectively, flag 12 is not used in
3XMM. It should be noted that bright sources that suffer significant 
pile-up are not flagged in any way in 3XMM (or in previous {\em XMM-Newton} 
X-ray source catalogues).

The masked area of each image is an indicator of the quality of the field as a
whole. Large masked areas are typically associated with diffuse extended
emission, very bright sources whose wings extend across much of the image, or
problems such as arcs arising from single-reflected X-rays from bright sources
just outside the field of view. The fraction of the field of view that is
masked is characterised by the observation class ({\it OBS\_CLASS}) parameter. The
distribution of the six observation classes in the 3XMM catalogues has changed
with respect to 2XMMi-DR3 (see table~\ref{tab:obsclasses}). The dominant 
change is in the split of fields assigned observation classes 0 and 1, with
more fields that were deemed completely clean in 2XMMi-DR3 having very small
areas (generally single detections) being marked as suspect in the 3XMM
catalogues. Often these are features that were considered, potentially, to be 
unrecognised bright pixels, e.g. detections dominated by a single bright pixel
in one instrument with no similar feature in the other instruments. It should
be emphasised, however, that the manual screening process is unavoidably 
subjective.

\begin{table}[ht]
\begin{minipage}[ht]{\columnwidth}
\normalsize
\caption{3XMM observation classification ({\it OBS\_CLASS}) (first column),
  percentage of the field considered problematic (second column) and the percentage of fields 
that fall within each class for 2XMMi-DR3 and 3XMM-DR5 (third and fourth columns respectively)}
\label{tab:obsclasses}
\small
\centering
\renewcommand{\footnoterule}{} 
\tabcolsep 0mm
\begin{tabular}{
c @{\extracolsep{1mm}} c @{\extracolsep{1mm}} c @{\extracolsep{1mm}} 
c }
\hline \hline
OBS CLASS & masked fraction & \multicolumn{1}{c}{2XMMi-DR3} &
\multicolumn{1}{c}{3XMM-DR5}  \\
\hline
0 &  bad area = 0\%               & 38\%  & 27\% \\
1 &  0\% $<$ bad area $<$ 0.1\%   & 12\%  & 22\% \\
2 &  0.1\% $<$ bad area $<$ 1\%   & 10\%  & 12\% \\
3 &  1\% $<$ bad area $<$ 10\%    & 25\%  & 24\% \\
4 &  10\% $<$ bad area $<$ 100\%  & 10\%  & 11\% \\
5 &  bad area = 100\%             &  5\%  &  4\% \\

\hline
\end{tabular}
\end{minipage}
\normalsize
\end{table}

\section{Catalogue construction: unique sources}\label{sec:cat_contruct}
The 3XMM detection catalogues collate all individual detections from the 
accepted observations. Nevertheless,
since some fields have at least partial overlaps with others and some targets
have been observed repeatedly with the target near the centre of the field of
view, many X-ray sources on the sky were detected more than once (up to 48 
times in the most extreme case). Individual detections have been assigned to
unique sources on the sky (i.e. a common unique source identifier, SRCID, has
been allocated to detections that are considered to be associated with the
same unique source) using the procedure outlined here. The process used in
constructing the 3XMM catalogues has changed from that used for the 2XMM 
series of catalogues (\cite{wats09} - see their section 8.1).

The matching process is divided into two stages. The first stage finds, for
each detection, all other matching detections within 15\arcsec\ of it, from
other fields (i.e. excluding detections from within the same observation,
which, by definition, are regarded as arising from distinct sources) and
computes a Bayesian match probability for each pair as \cite{BS2008}

\begin{equation}
p_{match} = \Bigg[ 1 + {\frac{1-p_0}{B\cdot{p_0}} }\Bigg]^{-1}
\end{equation}

Here, $B$, the Bayes factor, is given by

\begin{equation}
B = {\frac{2}{(\sigma^2_1 + \sigma^2_2)} } \exp-\Bigg[\frac{\psi^2}
  {2(\sigma^2_1 + \sigma^2_2)} \Bigg]
\end{equation}
  
where $\sigma_1$ and $\sigma_2$ are the positional error radii of each
detection in the
pair (in radians) and $\psi$ is the angular separation between them, in
radians.  $p_0 = N_* / N_1 N_2$ where $N_1$ and $N_2$ are the numbers of
objects in the sky based on the surface densities in the two fields and $N_*$ is
the number of objects common between them. Each of these $N$ values is 
derived from the numbers of detections in the two fields and are then scaled
to the whole sky. The value of $N_*$ is not known, a~priori, and in general
can be obtained iteratively by running the matching algorithm. However, here
we are matching observations of the same field taken with the same telescope
at two different epochs so that in most cases, objects will be common. Of
course this assumption is affected by the fact that the two observations being
considered may involve different exposure times, different instruments,
filters and modes used and different boresight positions (with sources within
their fields of view being subject to different vignetting factors). To gauge
the impact of such effects in determining $N_*$, trials using an iterative
scheme were run, which indicated that taking $N_* = 0.9 min(N_1,N_2)$ provides
a good estimate of $N_*$ without the need for iteration. Finally, with all
pairs identified and probabilities assigned, pairs with $p_{match} < 0.5$ were
discarded.

In the second, clustering stage, a figure-of-merit is computed for each
detection, referred to here as the goodness-of-clustering (GoC), which is the
number of matches the detection has with other detections, normalised by the
area of its error circle radius (given by POSERR, see Appendix~\ref{ap:defs}). This GoC measure
prioritises detections that lie towards the centre of a group of detections,
and are thus likely to be most reliably associated with a given unique
source. The list of all detections is then sorted by this GoC value. The
algorithm works down the GoC-sorted list and for each detection, the other
detections it forms pairs with are sorted by $p_{match}$. Then, descending
this list of pairs, for each one there are four possibilities for assigning
the unique source identifiers: i) if both detections have previously been
allocated to a unique source and already have the same SRCID, nothing is done,
ii) if neither have a SRCID, both are allocated the same, new SRCID, iii)
if only one of them has already been assigned a SRCID, the other is
allocated the same SRCID, iv) where both detections in the pair have
allocated but different SRCIDs, this represents an ambiguous case - for
these, the existing SRCIDs are left unchanged but a confusion flag is set
for both detections.

This approach is reliable in matching detections into unique sources in the
large majority of cases. Nevertheless, there are situations where the process
can fail or yield ambiguous results. Examples typically arise in
complex regions, such as where spurious sources, associated with diffuse
X-ray emission or bright sources, are detected and, by chance, are spatially
close to the positions of other detections (real or spurious) in other
observations of the same sky region. Often, in such cases, the detections
involved will have manual quality flags set (\cite{wats09} - see their section
7.5 and also section~\ref{sec:screening} above.

Other scenarios that can produce similar problems include i) poorly
centroided sources, e.g. those suffering from pile-up or optical loading,
ii) cases where frame rectification (see~\ref{sec:rectification}) fails and
positional uncertainties are larger than the default frame-shift error of
1.5\arcsec\ that is adopted for un-rectified fields, iii) sources
associated with artefacts such as out-of-time event features arising from
bright objects elsewhere in the particular CCD, or residual bright pixels and
iv) where multiple detections of sources that show notable proper motion 
(which is not accounted for in pipeline processing) can end up being grouped
into more than one unique source along the proper motion vector. Overall, 
in 3XMM-DR5, this matching process has associated 239505 detections
with 70453 unique sources that comprise more than one detection.

\section{External catalogue cross-correlation}\label{sec:catxcorr}
The {\em XMM-Newton} pipeline includes a specific module, the Astronomical Catalogue Data Subsystem (ACDS), running at the {\em Observatoire de Strasbourg}. This module lists possible multi-wavelength identifications and generates optical finding charts for all EPIC detections. Information on the astrophysical content of the EPIC field of view is also provided by the ACDS. 

When possible, finding charts are built using $g$-, $r$- and $i$-band images extracted from the SDSS image server and assembled in false colours. Outside of the SDSS footprint, images are extracted from the Aladin image server.   
The list of archival astronomical catalogues used during the 3XMM processing includes updated versions of those used for the 2XMM and adds some of the most relevant catalogues published since 2007. A total of 228 catalogues were queried including Simbad and NED. Note that NED entries already included in ACDS catalogues (e.g. SDSS) were discarded.   
 
Among the most important additions are: \\
i) the Chandra source catalogue version 1.1 \citep{evan10}. This release contains point and compact source data extracted from HRC images as well as available ACIS data public at the end of 2009. ACDS accesses the Chandra source catalogue using the VO cone search protocol, \\
ii) the Chandra ACIS survey in 383 nearby galaxies \citep{liu11}, \\
iii) the SDSS Photometric Catalog, Release 8 \citep{aiha11}, \\
iv) the MaxBCG galaxy clusters catalogue from SDSS \citep{koes07}, \\
v) the 2XMMi/SDSS DR7 cross-correlation \citep{pine11}, \\
vi) the 3rd release of the RAVE catalogue \citep{sieb11}, \\
vii) the IPHAS H$\alpha$ emission line source catalogue \citep{with08}, \\
viii) the WISE All-Sky data Release \citep{cutr12}, \\
ix) the AKARI mid-IR all-sky survey \citep{ishi10} and version 1.0 of the all-sky survey bright source catalogue \citep{yama10}, \\
x) the Spitzer IRAC survey of the galactic center \citep{rami08}, \\
xi) the GLIMPSE Source Catalogue \citep[I + II + 3D][]{chur09}, \\
xii) the IRAC-24micron optical source catalogue \citep{sura04} \\
and xiii) the Planck Early Release Compact Source Catalogue \citep{planckVII}. 

Table \ref{xmatchstat} lists, for a selection of archival catalogues, the number of EPIC detections having at least one catalogue entry in the 99.73\% (3 Gaussian $\sigma$) confidence region.

\begin{table}
\begin{center}
\caption{Cross-matching statistics between 3XMM sources and other catalogues.}
\label{xmatchstat}
\begin{tabular}{lrlr}
\hline
Catalogue & Detections & Catalogue & Detections\\
\hline
Chandra src cat.& 63,676 & Chandra gal. & 9,908 \\
SDSS8 & 129,252 & RAVE & 219\\
USNO-B1.0 & 229,730 & IPHAS &38\\
WISE & 454,957 & AKARI &5,598\\
2MASS & 36,830 & GLIMPSE & 35,572\\
Simbad & 204,657 & Planck ERCSC & 43,136\\
NED & 296,914  \\
\hline
\end{tabular}
\end{center}              
\end{table}

The cross-matching method used for 3XMM is identical to that applied in
the former XMM catalogues. Briefly speaking, ACDS retains all archival
catalogue entries located within the 99.73\% confidence region around the position of the EPIC detection. The corresponding error ellipse takes into account systematic and statistical uncertainties on the positions of both EPIC and archival catalogue entries. The 3XMM implementation of the ACDS assumes that the error distribution of EPIC positions is represented by the 2-D Gaussian distribution 
\[ f(\delta_{RA},\delta_{DEC})\ =\ \frac{1}{2\pi\sigma_{RA} \sigma_{DEC}}\, e^{-\frac{1}{2}\,(\frac{\delta_{RA}^{2}}{\sigma_{RA}^{2}}+\frac{\delta_{DEC}^{2}}{\sigma_{DEC}^{2})})} \]
with 
\[
\sigma_{RA}\,=\,\sigma_{DEC}\,= \sqrt{\mathrm{RADEC\_ERR}^{2}/2+\mathrm{SYSERRCC}^{2}/2}
\]
thereby correcting for the overestimated error value used during the 2XMM processing.

ACDS identifications are not part of the 3XMM catalogue fits file but are made
available to the community through the XSA and through the
XCAT-DB\footnote{http://xcatdb.unistra.fr}, a dedicated interface developed
by the SSC in Strasbourg \citep[Michel et al. in press]{motc09}. The XCAT-DB also gives access to the entire 3XMM catalogue and to some of the associated pipeline products such as time series and spectra. Quick look facilities and advanced selection and extraction methods are complemented by simple X-ray spectral fitting tools. In the near future, the database will be enriched by the multi-wavelength statistical identifications and associated spectral energy distributions computed within the ARCHES project \citep[][]{motc14}.  Spectral fitting results from the XMMFITCAT database \citep{corr15} are also partially available.

\section{Known problems in the catalogue} \label{sec:knownprobs}
A number of small but significant issues have been identified that affect the
data in the 3XMM catalogues. Two of these affect both the 3XMM-DR4 and 
3XMM-DR5 catalogue. The other issues affect only 3XMM-DR4 and are described 
in appendix~\ref{ap:knownissues}.
  
\begin{enumerate}

\item The optimised flare filtering process (see section~\ref{sec:optflare}) returns a
  background rate threshold for screening out background flares for each
  exposure during processing. However, while this process generally works
  well, when the background level is persistently high throughout the
  observation, the optimised cut level, while formally valid, can still result
  in image data with a high background level. In principle, such cases could
  be identified by testing the cut threshold against a pre-determined
  benchmark for each instrument. However, this is complicated by the fact
  that, since the analysis is now measured in-band, apparently high
  background levels can also arise in fields containing large extended
  sources. To simplify the process of identifying affected fields, a visual
  check was performed during manual screening and fields where high background
  levels were suspected were noted and detections from those fields are
  flagged in the 3XMM catalogues via the {\it HIGH\_BACKGROUND} column.
  This screening approach has been somewhat conservative and subjective. A
  total of 21779 (20625) detections from 568 (552) {\em XMM-Newton} 
  observations are flagged as such in the 3XMM-DR5: numbers in parentheses are
  for 3XMM-DR4.

\item A further issue recognised in the 3XMM catalogue is that of 
detections in the previous 2XMMi-DR3 catalogue that are not present in the 
3XMM catalogue. There are 4921 {\em XMM-Newton} observations that are common
between 2XMMi-DR3 and the  3XMM-DR5 catalogues, resulting in 349444 detections in 2XMMi-DR3 and 359505 detections in 3XMM-DR5. Of these, there are 274564 point-like detections with a sumflag$\leq$1 in 2XMMi-DR3 and 283436 in 3XMM-DR5.  However, amongst these
observations, there are $\sim$54000 detections that appear in 2XMMi-DR3 that
are not matched with a detection in the same observation in the 3XMM-DR5 
catalogue within 10\arcsec. About 25700 of these were classified as the
cleanest ({\it SUM\_FLAG}$\le1$), point-like  sources in 2XMMi-DR3 - these are
referred to as missing 3XMM detections in  what follows. It should be noted
that in reverse, there are $\sim$64000 detections in the 3XMM catalogues that
are in common observations but not matched with a detection in 2XMMi-DR3
within 10\arcsec, approximately 33600 of which are classed as being clean 
and point-like. 

The details explaining these 'missing sources' are given in Appendix~\ref{ap:miss_srcs}, but the main reasons for the source discrepancies between the two catalogues are two of the major improvements to the 3XMM catalogue with
respect to the 2XMM catalogue, namely the new empirical PSF, described in
Sec.\ref{sec:PSF} and the optimised flare filtering (see
Sec.\ref{sec:optflare}). Other origins, such as MIP events which were present in 2XMMi-DR3 but not recognised as such, and mostly removed in 3XMM-DR5, also contribute, but to a lesser extent, to the missing sources. In general though, the improvements to the pipeline that was used to create the 3XMM-DR5 data introduce (generally small) changes in the likelihood values (of mostly real sources). As such, the imposed threshold cut at L=6 for inclusion in the catalogue results in a fraction of sources that had L$<$6 in 2XMMi-DR3 having a likelihood a little above it in the 3XMM-DR5 processing, and vice versa, leading to losses and gains between the catalogue. Overall, the processing for 3XMM-DR5 is shown to be an improvement over the 2XMMi-DR3 procedure (see Appendix~\ref{ap:miss_srcs}), resulting in more sources.

\end{enumerate}

\section{Catalogue characterisation}\label{sec:CATCHAR}

\subsection{General properties}\label{sec:CATCHARgen}
\begin{figure}
   \centering
   \includegraphics[width=9cm]{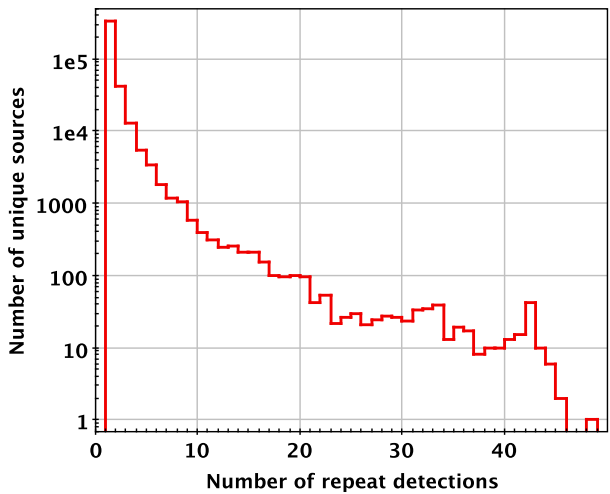}
      \caption{Numbers of 3XMM-DR5 unique sources comprising given numbers of
        repeat detections.}
         \label{fig:repeat_dets}
   \end{figure}

The 3XMM-DR5 catalogue contains 565962 (531261) detections, associated with
396910 (372728) unique sources on the sky, extracted from 7781 (7427) public 
{\em XMM-Newton} observations - numbers in parentheses are for
3XMM-DR4. Amongst these, 70453 (66728) unique sources have multiple
detections, the maximum number of repeat
detections being 48 (44 for 3XMM-DR4), see fig.~\ref{fig:repeat_dets}. 55640
X-ray detections in 3XMM-DR5 are identified as extended objects, i.e. with a
core radius parameter,
$r_{core}$, as defined in section 4.4.4 of \cite{wats09}, > 6\arcsec, with
52493 of these having $r_{core} <$~80\arcsec.  Overall properties in terms of
completeness and false detection rates are not expected to differ
significantly from those described in \cite{wats09}.

\subsection{Astrometric properties}\label{sec:CATCHARastrometry}

As outlined in section~\ref{sec:rectification}, several changes have been made
to the processing that affect the astrometry of the 3XMM catalogues relative to 
previous {\em XMM-Newton} X-ray source catalogues. To assess the quality of 
the current astrometry, we have broadly followed the approach outlined in
\cite{wats09}. Detections in the 3XMM-DR5 catalogue were cross-correlated
against the SDSS DR12Q quasar catalogue (Paris et
al. in prep.), which contains $\sim$297300 objects spectroscopically
classified as quasars - positions and errors were taken from
the SDSS DR9 catalogue. X-ray detections with an SDSS quasar counterpart within
15\arcsec\ were extracted. Point-like 3XMM-DR5 detections were selected with 
summary flag 0 (see Appendix~\ref{ap:defs}), from successfully {\it catcorr}-corrected fields, with EPIC 
detection log-likelihood $>$8 and at off-axis angles < 13\arcmin. The SDSS quasars 
were required to have warning flag 0, morphology 0 (point-like) and r' and g'
magnitudes both $<$22.0.  This yielded a total of 6614 3XMM-QSO pairs. In the
13 cases where more than one optical quasar match was found within 15\arcsec,
the nearest match was retained. 

The cross-matching used the {\it catcorr}-corrected {\it RA} and {\it DEC} X-ray
detection coordinates. The measured separation, $\Delta{r}$, and the 
overall 1-dimensional XMM position error, $\sigma_{1D}$ ($=
\sigma_{pos}/\sqrt2$), were recorded. Here $\sigma_{pos}$ is the radial
positional error, {\it POSERR}, in the catalogues, which is the quadrature sum 
of the XMM positional uncertainties resolved in the RA and DEC directions. 
As noted by \cite{wats09}, if the offset between the X-ray source and
its SDSS quasar counterpart, $\Delta{r}$, is normalised by the total position 
error, $\sigma_{tot}$, i.e. $x = \Delta{r} / \sigma_{tot}$, the distribution
of these error-normalised offsets is expected to follow the Rayleigh 
distribution, 

\begin{equation}\label{eq:rayleigh}
N(x)dx \propto {x e^{-x^2 /2} dx} 
\end{equation}

Errors on the SDSS quasar positions were included in $\sigma_{tot}$ though
they are generally $\le$ 0.1\arcsec, much smaller than 
the vast majority of $\sigma_{1D}$ values in 3XMM-DR5. The SDSS position
errors were circularised using $\sigma_{QSO} = [ (\sigma^2_{maj} +
\sigma^2_{min})/2]^\frac{1}{2}$ where $\sigma_{maj}$ and  $\sigma_{min}$ are
the errors in the major and minor axis directions of the SDSS position error
ellipse. These were then combined in quadrature with the XMM position error 
to obtain $ \sigma_{tot} = (\sigma^2_{1D} +\sigma^2_{QSO})^\frac{1}{2}$.
No systematic error was included for the QSO position error.

\begin{figure}
   \centering

   \includegraphics[height=7cm,width=9cm]{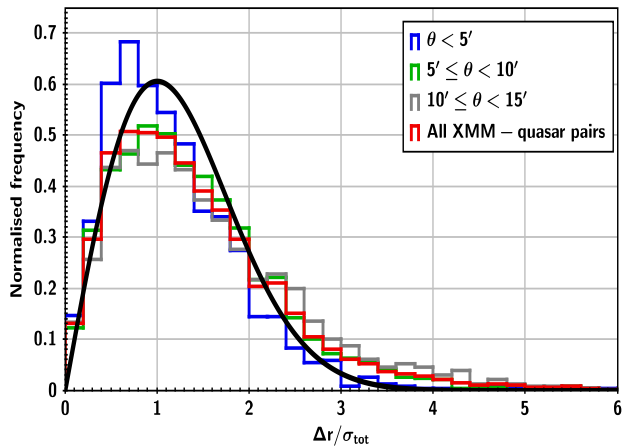}

      \caption{Distribution of position-error-normalised offsets between 
3XMM-DR5 X-ray
        sources and SDSS quasar counterparts (red histogram). The expected
        Rayleigh distribution is overlaid (black). The XMM position errors are
        as provided in the 3XMM catalogues (i.e. unadjusted, with no scaling or
        systematic included).  Also shown are similar histograms for data from
      EPIC off-axis angles, $\theta$,  in the ranges $\theta<5'$ (blue), $5'\le\theta< 10'$
      (green) and $10'\le\theta< 15'$ (grey). The data are normalised to unit area. }

         \label{fig:rayleigh_corr}
   \end{figure}

In Fig.~\ref{fig:rayleigh_corr} we show the distribution of $x$ values for the
selected XMM-QSO pairs as the red histogram, with the expected Rayleigh 
distribution overlaid in black. While there is broad overall agreement 
between the data and model, it is clear that there is a deficit of sources 
for $0.8 < x < 2$ and an excess for $x > 2.5$. A total of 739 XMM-QSO pairs 
lie at $2.5 <x < 6$ while the model predicts 291, the excess of 448 
representing 6.8\% of the total in the histogram.

To investigate the small discrepancies between the distribution of $x$ values for the
selected XMM-QSO pairs and the Rayleigh distribution, we carried out a number of tests detailed in Appendix~\ref{ap:ast_rayleigh}. The main results of these tests indicate that the excess of 3XMM-DR5 detections with error-normalised
offsets from their SDSS quasar counterparts $> 3.5$ appears to have a modest 
dependence on the off-axis location of the detection in the XMM field of view. A small fraction of detections at higher off-axis angles have either 
incorrect positions or underestimated errors, while sources near the centre 
may have slightly overestimated errors. Further, given that the sources at higher off-axis angles in EPIC images have rather elongated PSF profiles, the assumption of a circularly symmetric positional uncertainty distribution is probably not adequately representative of such sources, which may be contributing to the observed discrepancies.

\begin{figure}
   \centering

   \includegraphics[height=8cm,width=7cm, angle=270]{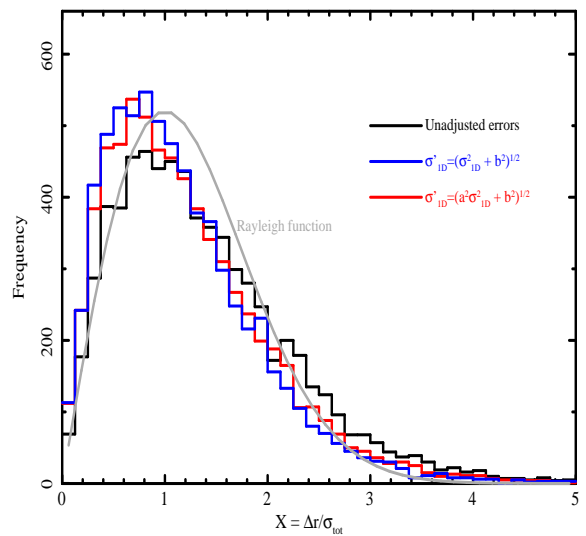}

      \caption{Similar to figure~\ref{fig:rayleigh_corr} but comparing results 
        that involve the simplest adjustments to the XMM position errors. For 
        reference, the black histogram is based on using the unadjusted XMM 
        position errors while the expected
        Rayleigh distribution is overlaid (grey). The blue histogram
        represents the simplest adjustment to the XMM position errors, involving the
        addition of a systematic in quadrature, $b$(=0.37), while the red histogram involves both
        a scaling of the XMM position error by a factor $a$(=1.12) and
        addition of a systematic, $b$(=0.27), in quadrature. These histograms are based
        on slightly different filtering compared to
        figure~\ref{fig:rayleigh_corr}, as explained in the text}
         \label{fig:rayleigh_tests}
   \end{figure}

Subsequent analysis investigated whether the discrepancies could be
reduced by making phenomenological adjustments to the XMM position errors. 
In this analysis, the filtering applied to XMM and SDSS sources was similar to 
that outlined above but only matches within 5\arcsec\ were used and no
magnitude limits were imposed on the SDSS objects, resulting in 6858 pairs. 
A two parameter adjustment was considered in which the XMM position errors
were scaled by a constant, $a$, and a systematic error, $b$, was added in
quadrature (i.e. $\sigma'_{1D} = (a^2\sigma^2_{1D} +
b^2)^\frac{1}{2}$). One parameter adjustments, where only the systematic was
added (i.e. where $a$ is set to 1) were also tested. 
The error normalised XMM-quasar separations were recomputed as
$x' = \Delta{r} / \sigma'_{tot}$,  where $\sigma'_{tot}$ now combines
$\sigma'_{1D}$ with  $\sigma_{QSO}$ in quadrature. Using this prescription,
the data were fit to the Rayleigh function to obtain the best-fit values for
$a$ and/or $b$, using a maximum likelihood approach. The results are shown in
figure~\ref{fig:rayleigh_tests}. While these parameterisations of the XMM
position errors did improve the fit, particularly bringing the data in the
tail closer to the expected 
Rayleigh curve, the fit remains poor overall, driving the peak of the data to 
$x \approx 0.7$ (it should peak at 1.0) and introducing a notable excess at 
$x < 1$. Despite the fact these two forms of adjustment to the XMM position 
errors yield statistically unacceptable fits to the Rayleigh function, as they do improve agreement 
in the tail (i.e. for a given XMM-counterpart pair, $x' < x$), they reduce the
chance that real matches of 3XMM sources with counterparts from other
catalogues (or observations) will be erroneously excluded as candidate
counterparts. As such, although the position error column values in the 3XMM 
catalogues are not adjusted, we provide the values of
$a$(=1.12) and $b$(=0.27\arcsec) for the two parameter fit so that users can
apply the above adjustments to the XMM position errors if they wish. The one 
parameter case best fit yields $b=0.37$.

Other tests 
involved (i) imposing a lower bound on the XMM position error ($\sigma'_{1D} =
max(\sigma_{1D} , \sigma_{min}) )$ and (ii) including an off-axis-dependent 
systematic involving a scalar, $c$,  
$(\sigma'^2_{1D} = \sigma^2_{1D} + c^2 {\Theta^2})$ where $\Theta$ is 
the off-axis angle. These latter modifications provide
slightly better matches to the Rayleigh curve but still drive the peak of 
the data to $x \approx 0.7$, again creating an excess at $x < 1$.  A further 
test in which the XMM position error is defined as $\sigma'_{1D} =
\sigma_{1D}$ for $x < x_t$ and $\sigma'^2_{1D} = \sigma^2_{1D} + d^2 (x-x_t)^2$
for $x \ge x_t$ (where $d$ is a simple scalar and $x_t$ is a threshold value
in $x$) does yield a marked enhancement in the likelihood for the fit
but in this case, the data undershoot the Rayleigh curve at $x > 2$ and exceed
it at $0.6 < x < 2$.  

We conclude that while the more complex adjustments to the XMM position
errors can formally improve the match between the error-normalised XMM-quasar 
separations and the Rayleigh curve, none provides a statistically acceptable 
match. Moreover, the cases that yield the best improvements in the fit
likelihood have no compelling technical rationale.

\subsection{Background flare filtering }\label{sec:CATCHARoptflare}

As noted in section~\ref{sec:optflare}, an optimisation algorithm was 
adopted to determine the count rate threshold for defining the flare GTIs. 
This process was employed to maximise sensitivity to source detection and can
come at the expense of reduced exposure time. Often, the new process results
in GTIs that are similar to those derived from the fixed threshold cuts used
in pre-cat9.0 pipeline processing. However, in some cases, significant
improvements can be obtained in sensitivity.

\begin{figure}
   \centering
   \includegraphics[height=9cm, width=4.cm,
     angle=270]{0049350101M1S001FBKTSR_2xmmidr3_v_3xmm_060115.eps}
   \includegraphics[width=9cm]{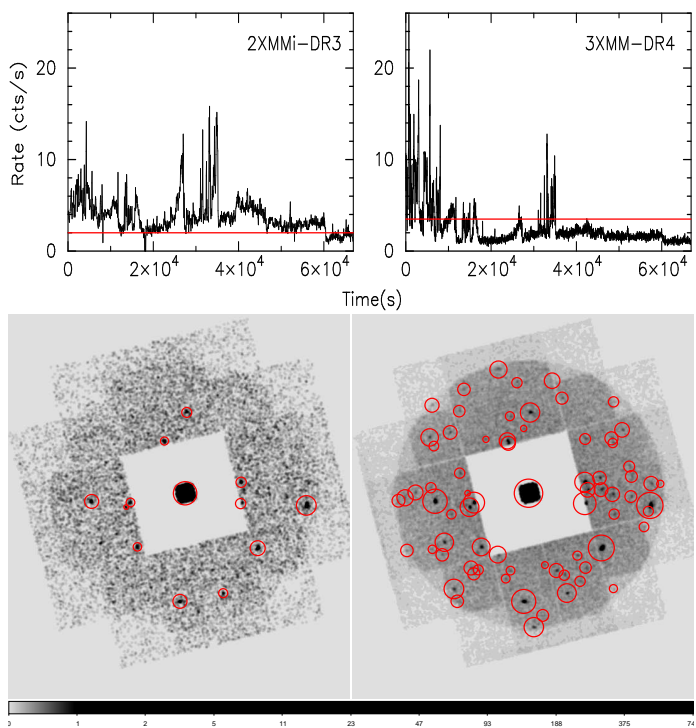}

      \caption{An example of the improvement offered by the optimised
        background flare filtering algorithm. Top panels: Left: high-energy
        MOS1 background flare light curve created by the pre-cat9.0
        pipeline, used for the 2XMMi-DR3 catalogue - the red line is the fixed
        (2 cts/s/arcmin$^2$) count rate cut threshold applied. Right: in-band
        (0.5-7.5~keV) light curve used in the cat-9.0 pipeline used for
        3XMM-DR4 and 3XMM-DR5 - the red
        line shows the optimised cut rate threshold derived for the light
        curve. The lower panels show the resulting, corresponding (smoothed) images,
        after filtering out the data above the respective rate-cut
        thresholds. Sources found by the source detection algorithm are
        indicated by red circles.}
         \label{fig:goodflareexample}
   \end{figure}

Of particular interest are cases where the background rises or falls
slowly. In such cases, allowing a modest increase in the background count rate  
can yield a marked increase in exposure time, resulting in a significant 
improvement in the sensitivity to the detection of faint sources. A good
example of this is illustrated in figure~\ref{fig:goodflareexample}.  As is
evident from the light curves, the optimised cut threshold includes
significantly more exposure time for a very modest increase in background
level, producing a factor 5.5 increase in the harvest of detected sources.

Another aspect of the optimised flare filtering approach is that the increase
in exposure time can result in exposures being used that were previously 
rejected in processing with pre-cat9.0 pipelines. 

The pre-cat9.0 and cat9.0 light curves in figure~\ref{fig:goodflareexample}
also highlight the fact that the change of energy band used can yield some
significant differences in the strengths and even shapes of flare features in 
the data.  

The implementation of the optimised flare filtering approach was done in
conjunction with some of the other upgrades, such as the use of the empirical
PSF (see section~\ref{sec:PSF}). As such, we have not directly
isolated the impacts on source detection of the optimised flare filtering
process alone.  Nevertheless, comparison of the numbers of source detections  
between the set of 4922 observations that are common to the 2XMMi-DR3 and 
3XMM-DR5 catalogues, indicates a net increase of 10047 detections in 3XMM-DR5, 
i.e. a 2.9\% increase.

\subsection{Extraction of spectral and time series products }\label{sec:CATCHARssp}

As described in section~\ref{sec:SSPoptext}, spectra and time series of
detections are now extracted using optimised extraction apertures that are
intended to maximise the overall S/N of the resulting product. To assess this, 
spectra were re-extracted for all detections and exposures
for which spectra were produced during the bulk reprocessing, using a circular
aperture of fixed radius (28\arcsec) in each case, centred at the same
location as the detection position used during the bulk reprocessing. Other 
than the change of aperture radius, processing was essentially identical to 
that used in the bulk reprocessing. The S/N, $S$, of each spectrum was then 
computed as $S = {C_s}/{{C_T}^\frac{1}{2} } $. Here $C_s = C_T - C_b$, where 
$C_T$ is the total number of counts measured in the spectrum from the source 
aperture, $C_s$ is the number of counts from the source in the source 
aperture and $C_b$ is the number of background counts in the source aperture, 
the latter being estimated from the total counts in the background
region, scaled by the ratio of source and background region areas. Counts
included in this analysis were drawn from PHA channels with quality $<5$ 
(in XSPEC terms). The S/N was computed in this way for the spectra 
from the optimised and fixed apertures - the spectral data used for background 
subtraction were taken from the same background spectrum (from the bulk 
reprocessing) in each case and the background counts used were drawn only 
from the same channels as used for the source counts. 

 \begin{figure}
   \centering
   \includegraphics[height=8cm, width=9.cm]{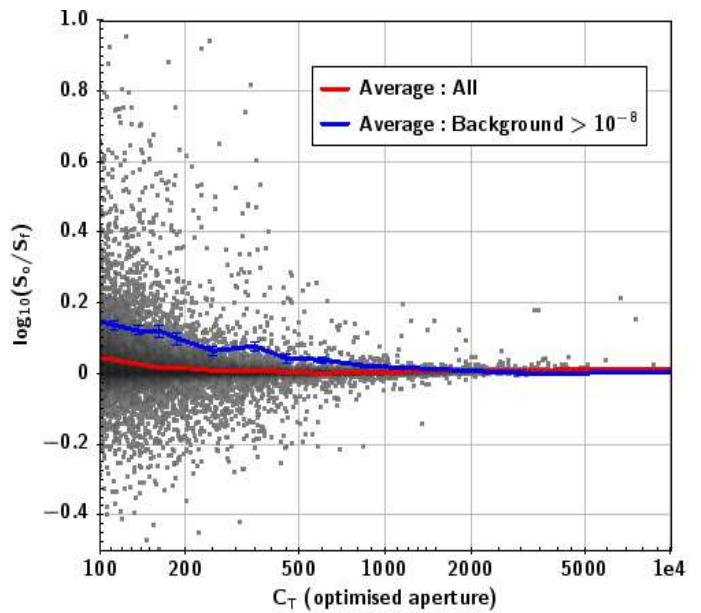}
   \caption{$log_{10}(S_o/S_f)$ plotted against the log of the
     total counts, $C_T$, measured from the optimised aperture. The grey 
     points indicate the data and include only clean ({\it SUM\_FLAG}=0), 
     point-like ({\it EP\_EXTENT} =0) detections. The red line links 
     measurements of the average $log_{10}(S_o/S_f)$, in bins
     sampling the range in $C_T$, for cases where 
     $-1 < log_{10}(S_o/S_f) < 1$. The
     blue line is similar but is for the subset of data where, additionally, 
     the background rate is $>10^{-8}$~cts~s$^{-1}$~(sub-pixel)$^{-2}$ 
(sub-pixels have side lengths of 0.05~\arcsec). The lower X-axis limit 
reflects the minimum threshold of 100 total counts in the optimised 
extraction aperture, imposed for extracting XMM source spectra; the plot is 
otherwise truncated for clarity.}
         \label{fig:sspsnfig}
   \end{figure}

In Figure~\ref{fig:sspsnfig} the log of the ratio of the S/N values from
the spectra extracted from the optimised ($S_o$) and fixed ($S_f$) apertures,
i.e. $log_{10}(S_o/S_f)$, is plotted against $\log(C_T)$ from the optimised 
aperture, for MOS1 spectra. Only spectra from the cleanest ({\it SUM\_FLAG}=0), 
point-like ({\it EP\_EXTENT}=0) detections are included. 

It is evident from the positive asymmetry about $log_{10}(S_o/S_f)=0$, that the optimisation procedure 
does improve the S/N of the spectra, especially for spectra with lower 
($C_T < 500$) numbers of extracted counts, as expected. Overall, 67.5\% of 
the MOS1 spectra with $100 < C_T < 50000$~cts (within $-1 < log_{10}(S_o/S_f)
<1$ which excludes 21 positive outliers) have higher S/N in the optimised aperture 
than those extracted from the fixed 
apertures.  The red line in figure~\ref{fig:sspsnfig} shows the average of 
$ log_{10}(S_o/S_f)$ of all the data as a function of $C_T$ and indicates
that spectra extracted from the optimised apertures with $C_T=100$~cts have,
on average, S/N values 12\% higher than those extracted in the fixed 
apertures. It is anticipated that sources detected in fields with
high background levels would benefit from the optimisation procedure. 
Indeed the blue line in figure~\ref{fig:sspsnfig}, which reflects the subset 
of detections whose background levels are above $10^{-8}$~cts
s$^{-1}$~(sub-pixel)$^{-2}$ (i.e. amongst the highest 15\% of background levels), 
demonstrates this - spectra of such detections 
extracted from optimised apertures with $C_T=100$~cts, have average S/N values 
~39\% higher than the spectra from the corresponding fixed apertures.

\subsection{Extended sources}
\label{sec:extended}

\begin{figure*}
\begin{center}
\includegraphics[width=0.45\textwidth]{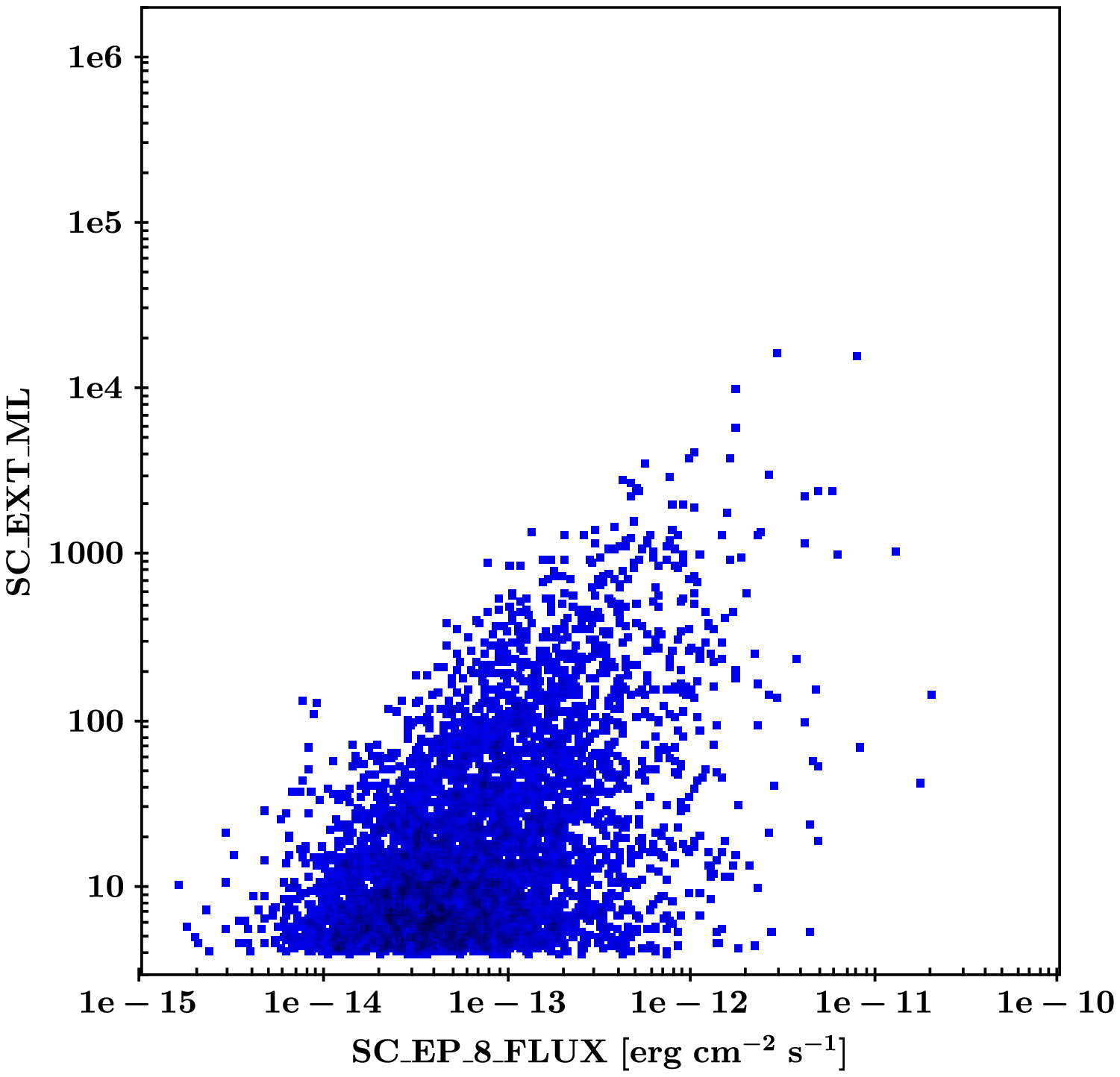}
\includegraphics[width=0.45\textwidth]{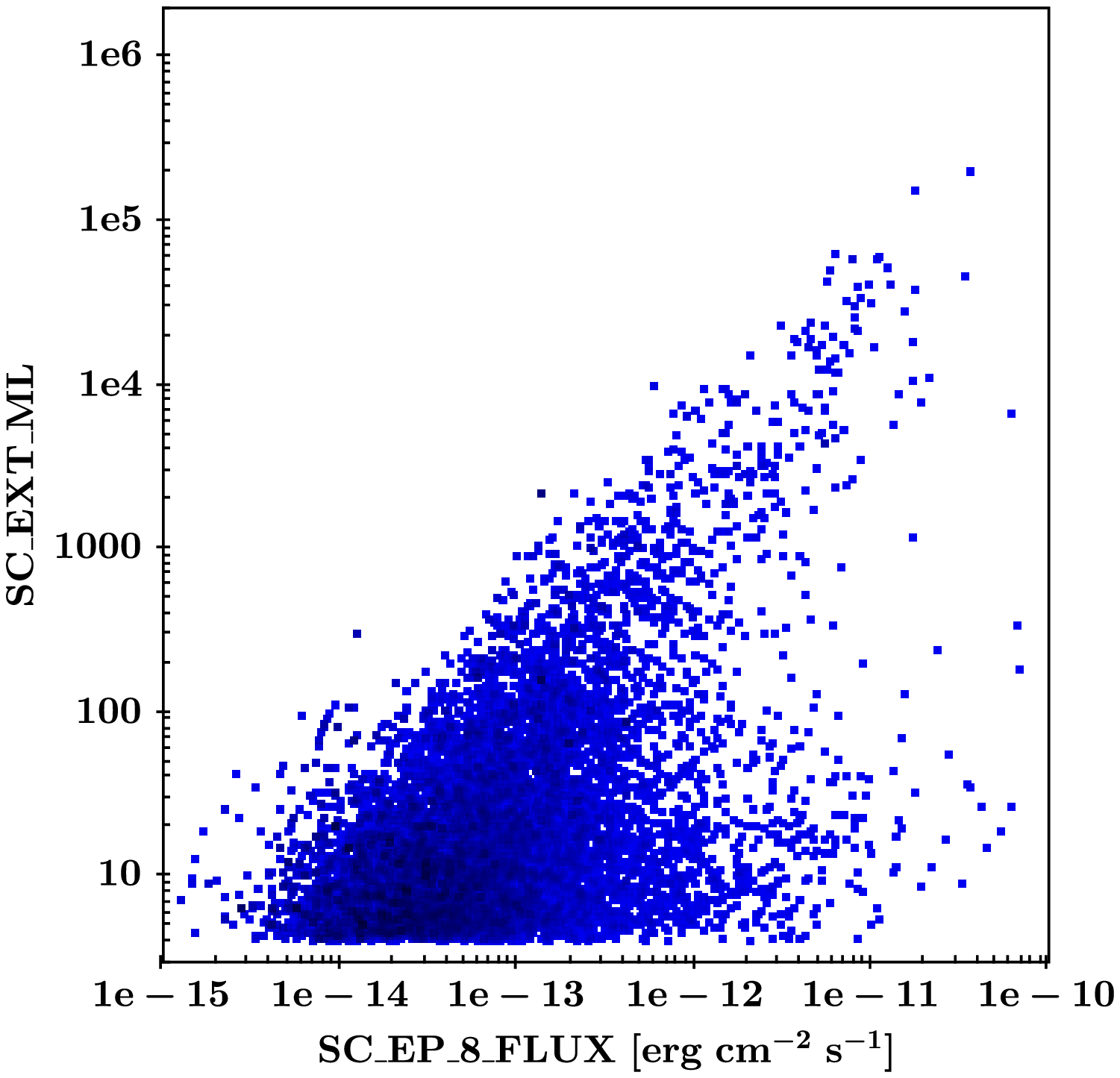}
\caption{Distribution of total flux and extension likelihood of all extended sources with SC\_SUM\_FLAG $<$ 2, in the 2XMM (left) and 3XMM (right) catalogues.}
\label{ext_3xmm_flag}
\end{center}
\end{figure*}

The detection and characterisation of extended sources for 3XMM was performed as in 2XMM \citep{wats09}. The caveats listed in section 9.9 of that paper still apply to 3XMM. However the better representation of the PSF has helped to improve  extended source detection and characterisation. Many extended sources with SC\_SUM\_FLAG = 4 in 2XMM now have SC\_SUM\_FLAG = 3 in 3XMM, indicating that the region is still complex but the detection itself is unlikely to be spurious. We have also looked at the distribution of extension likelihood vs. flux as in Fig.~15 of \citet{wats09}. Fig.~\ref{ext_3xmm_flag} shows that 3XMM considers many bright extended sources to be reliable (SC\_SUM\_FLAG $<$ 2) whereas in 2XMM most of them had higher flag values indicating more significant issues with the data quality.

\begin{figure}
\begin{center}
\includegraphics[width=0.5\textwidth]{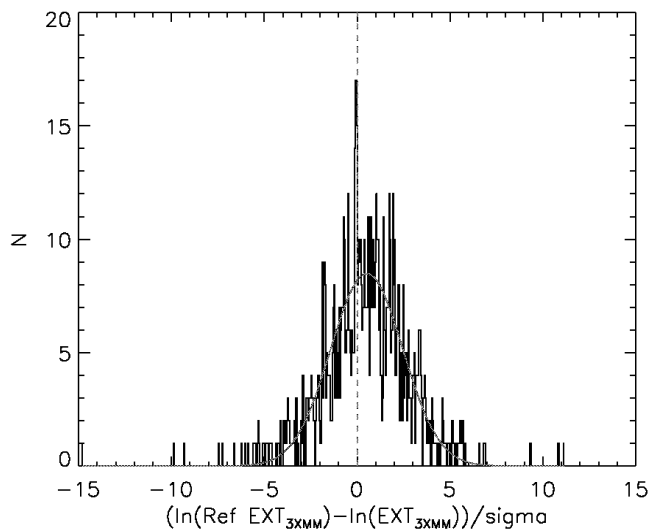}
\caption{Histogram of the logarithm of the ratio of extensions between the best observation and the other observations of the same source, normalised by the error.  The solid red  line is the best Gaussian fit to the histogram. The dashed red line is the expected mean ($0$).}
\label{ext_3xmm_his}
\end{center}
\end{figure}

We have complemented this study by inter-comparing the 3XMM (DR4) results when a source was observed more than once, and with an independent serendipitous search for clusters of galaxies. We restricted the comparison to the best-quality sources with SC\_SUM\_FLAG = 0. In 3XMM-DR4, 667 sources  have been observed several times as extended, each observation being processed independently. We define as the ``reference value'' the extension (EP\_EXTENT) associated to the detection with the highest likelihood value (EP\_8\_DET\_ML column). 
We investigated the agreement of the extension parameter between the ``reference" and the other observations of the same source. We ignored observations when a given source was not detected as extended (mostly because of insufficient exposure) or when the extension was set to $80\arcsec$ (maximum value allowed in the fit).

In Fig.~\ref{ext_3xmm_his} we show the distribution (in log space) of the ratio between the ``reference'' extension $Ext_{ref}$ and the current one $Ext_{cur}$, normalised by the corresponding error equal to  $\sqrt{(\sigma_{ref}/Ext_{ref})^2 + (\sigma_{cur}/Ext_{cur})^2}$, where $\sigma_{ref}$ and  $\sigma_{cur}$ are the extension errors for the ``reference'' and current observation respectively. We fit the histogram result by a Gaussian function, obtaining a mean value equal to 0.512 (in $\sigma$ units)  with a standard deviation equal to 1.943 (we would expect a mean of 0 and a standard deviation of 1 for random fluctuations).
We conclude that there exists an additional scatter larger than statistical (of unknown origin) and that the reference observation, which is also the deepest one, estimates a larger extension on average.

\begin{figure}
\begin{center}
\includegraphics[width=0.5\textwidth]{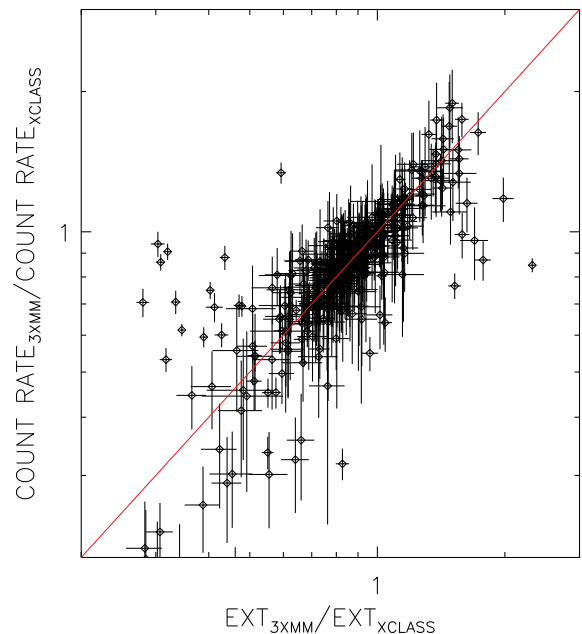}
\caption{Comparison of the ratio of extensions and the ratio of count rates obtained by the 3XMM  and the XCLASS catalogues. The red solid  line is the relation 1:1.}
\label{3xmm_xclass_ratio}
\end{center}
\end{figure}

The XCLASS catalogue is based on the analysis of archival observations from the XMM-Newton observatory. The XCLASS team processed 2774 high Galactic latitude observations from the XMM archive (as of 2010 May) and extracted a serendipitous catalogue of some 850 clusters of galaxies based on purely X-ray criteria, following the methodology developed for the XMM Large Scale Survey \citep{2007MNRAS.382..279P}. 
We used the subsample of 422 galaxy clusters available online at http://xmm-lss.in2p3.fr:8080/l4sdb/ to compare the extension and the count rate obtained for the same sources from the two different procedures (ie. the XCLASS and 3XMM processing). The analytic expression used to represent extended sources in XCLASS was the same as in 3XMM ($\beta$-model with $\beta$=2/3) so the numbers should be directly comparable. All 422 clusters are in 3XMM-DR4, but 59 (mostly faint or irregular objects) were classified as point sources.

For the 363  extended sources in common, we compared the extent and the count rate in the  [0.5-2.0] keV band obtained by 3XMM and XCLASS.
We found that, for both quantities, the 3XMM estimates seem to be biased low with respect to the XCLASS values.
The best fit regression on source extent resulted in a slope of 0.7 ($Ext_{3XMM} \simeq 0.7 Ext_{XCLASS}$).
Excluding clear outliers (difference of extension larger than $20\arcsec$, typically very faint sources or very bright sources affected by a strong pile-up) the slope increases to 0.85. 
We conclude that, even excluding these extreme sources, there remains a bias of $\simeq$ 15\% between the extensions estimated by 3XMM and XCLASS.

There exists a similar (a little smaller) bias on the count rate. 
However Fig.~\ref{3xmm_xclass_ratio} shows that there exists a close correlation between both ratios, implying that only one parameter describes the difference in extent and count rates and that, if the source extents were forced to agree, the count rates would agree too.
There is no obvious way to know whether the 3XMM or the XCLASS estimate is better but, together with the inter-3XMM comparison, this result indicates that the purely statistical extension error underestimates the real error.


\section{Examples}\label{sec:examples}
\begin{figure*}
   \centering
   \includegraphics[height=14cm, width=18cm]{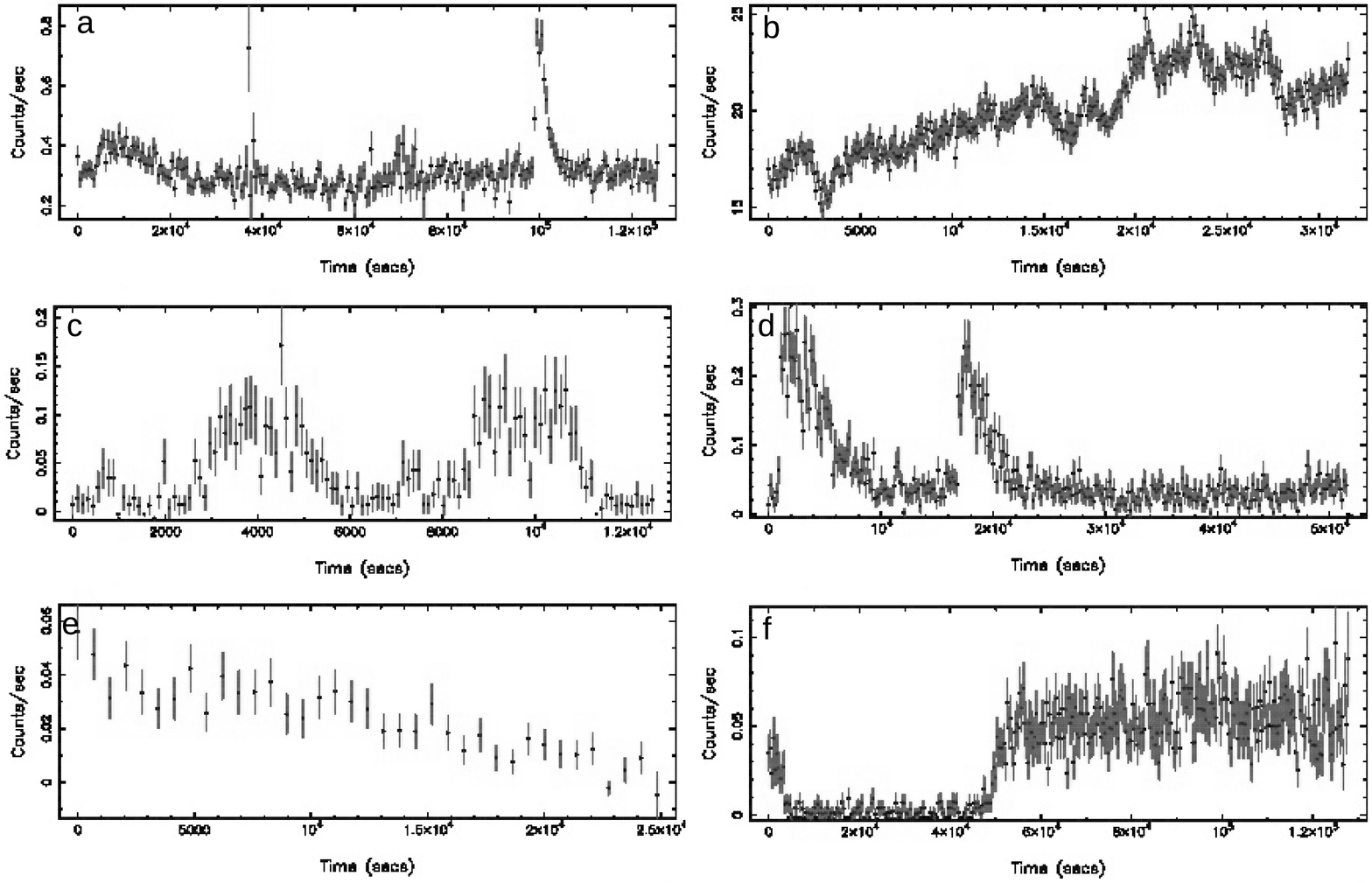}
      \caption{Example lightcurves taken directly from the 3XMM catalogue. a) 3XMM J111146.1-762010 = CHX 18N -- T Tau-type star showing a short flare b) 3XMM J000619.5+201210, A Seyfert 1, Markarian 335. c) 3XMM J184916.1+652943, a new 1.6 hour polar (Webb et al. to be submitted). d) A 2MASS star (2MASS J00025638-3004447) showing two large flares. e) 3XMM J002159.4+614254, a new X-ray object showing a strong decline in flux. Possibly a gamma-ray burst afterglow. f) 3XMM J013334.0+303211, a high mass X-ray binary in M 33, M33 X-7, showing a 12.5 hour eclipse - the first eclipsing stellar-mass black hole binary discovered  \citep{piet06}}
         \label{fig:example_lcs}
   \end{figure*}

\begin{figure*}
   \centering
   \includegraphics[height=14cm, width=18cm]{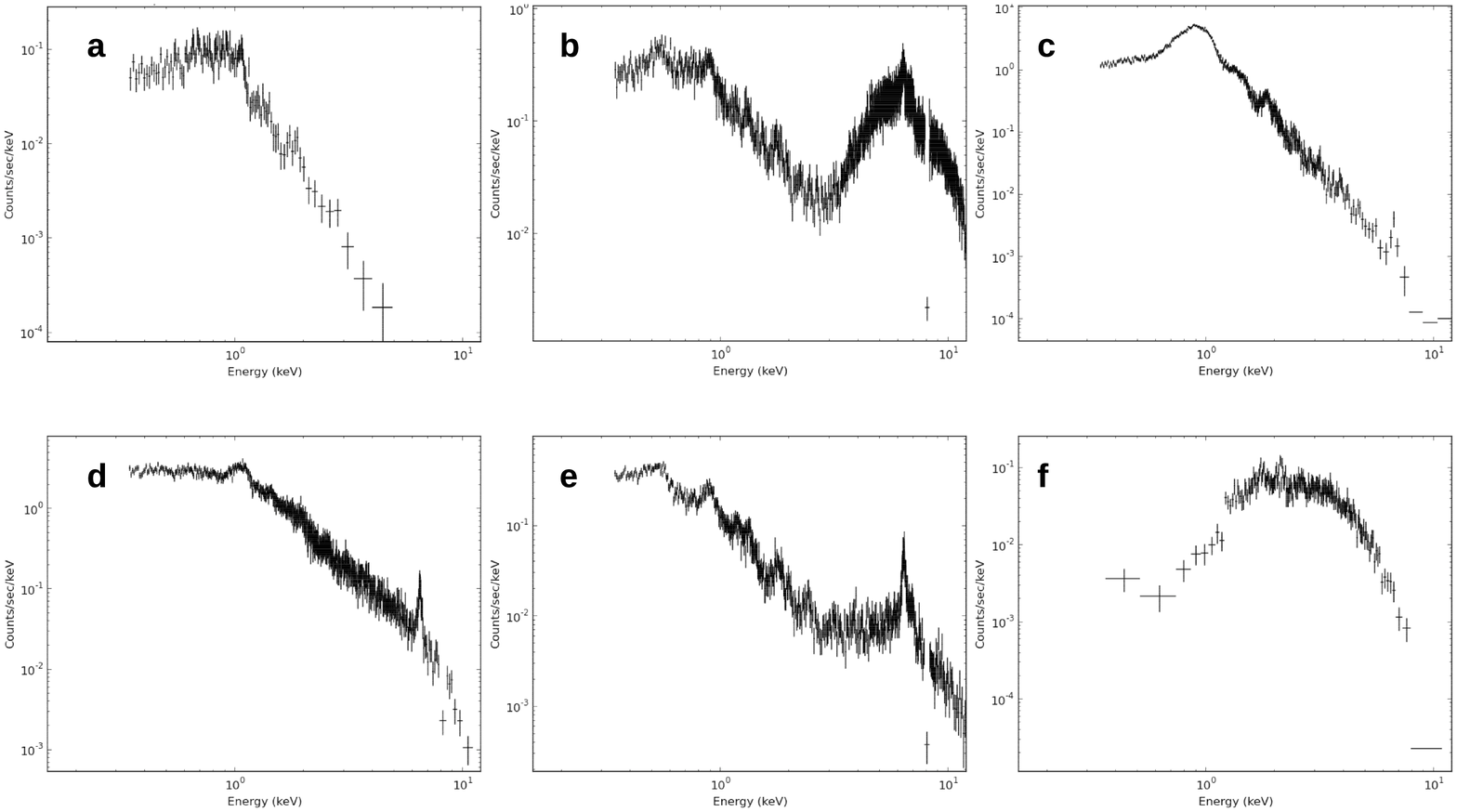}
      \caption{Example spectra taken directly from the 3XMM catalogue, showing the diversity of the sources in the 3XMM catalogue. Energy (keV) is given on the abscissa and count s$^{-1}$ keV$^{-1}$ on the ordinate. a) 3XMM J052532.5+062533, an X-ray source of unknown nature, as the majority of the sources are in 3XMM b) 3XMM J123536.6-395433 a Seyfert 2 galaxy (NGC 4507) c) 3XMM J125141.9+273226, a rotationally variable star, 31 Com  d) 3XMM J162838.2+393303, a cluster of galaxies e) 3XMM J011127.5-380500, the pn spectrum of NGC 424,  a Seyfert 2 galaxy f) 3XMM J185246.6+003317, a new transient magnetar discovered by \cite{zhou14}.  Some low data points can be seen in the plots originating from pn data, but these are corrected for when the spectra are plotted in conjunction with the distributed response files.}
         \label{fig:example_spectra}
   \end{figure*}

Thanks to the wide range of parameters  provided in the catalogue, sources matching specific criteria can be isolated (for  example variability criteria of X-ray hardness ratios). In this section we show  some examples of lightcurves (Fig.~\ref{fig:example_lcs}) and spectra (Fig.~\ref{fig:example_spectra}) extracted from the different EPIC cameras. The plots shown are those associated with the on-line catalogue. Both known and new sources are presented. It is immediately obvious from the two Figures that objects with extremely diverse characteristics are found. Variability on very different timescales is seen in Fig.~\ref{fig:example_lcs}, showing short and long flares, slow rises and steady declines in count rate as well as deep eclipses. From visual examination of the strong variability in Fig.~\ref{fig:example_lcs}c, it was quickly obvious that this new X-ray source was a polar (Webb et al. to be submitted). Fig.~\ref{fig:example_lcs}e shows a strong decline in flux, which, when coupled with the hard spectrum observed for this source, suggests that this might be a previously unknown orphan gamma-ray afterglow.

The spectra shown in Fig.~\ref{fig:example_spectra} are also very varied and originate from a variety of astrophysical objects, ranging from stars, compact objects, galaxies and clusters of galaxies. An unidentified X-ray source is included in Fig.~\ref{fig:example_spectra}a, which also has a highly variable lightcurve, showing a steady decline in count rate, but with a strong flare superposed. The nature of this source is not obvious and more work will be needed to identify its nature. The sources in the full 3XMM catalogue are of course dominated by unidentified objects, emphasising the large discovery space provided by the catalogue.

\section{Catalogue access}\label{sec:access}
The catalogue is provided in several formats. Firstly, a Flexible Image
Transport System (FITS) file and a comma-separated values (CSV) file is
provided containing all of the detections in the catalogue. For 3XMM-DR5 there
are 565962 rows and 323 columns. A separate version of the catalogue (the slim
catalogue) is also provided that contains only the unique sources, i.e. 396910
rows, and has 44 columns, essentially those containing information
about the unique sources. This catalogue is also provided in FITS and CSV
format. Ancillary tables to the catalogue also available from the
{\em XMM-Newton} Survey Science Centre
webpages\footnote{http://xmmssc.irap.omp.eu/} include the table of
observations incorporated in the catalogue and the target identification and 
classification table.

The {\em XMM-Newton} Survey Science Centre webpages provide access to the 3XMM
catalogue, as well as links to the different servers distributing the full
range of catalogue products. These include, the {\em XMM-Newton} XSA, which
provides access to all of the 3XMM data products, and the ODF data, the
XCat-DB\footnote{http://xcatdb.unistra.fr/3xmm/} produced and maintained by
the {\em XMM-Newton} SSC, which contains possible EPIC source identification produced 
by the pipeline by querying 228 archival catalogues. Finding charts are also provided 
for these possible identifications. Other source properties as well as images, time series, 
spectra, fit results from the XMMFITCAT are also provided. Multi-wavelength data 
taken as a part of the XID (X-ray identification project) run by the SSC over  the first
fifteen years of the mission are also provided in the XIDresult 
database\footnote{http://xcatdb.unistra.fr/xidresult/}. The LEDAS
server\footnote{http://www.ledas.ac.uk/arnie5/arnie5.php?action=basic\&catname=3xmm} 
provides another way to access the 3XMM catalogue and its products, whilst the upper limit
server\footnote{http://www.ledas.ac.uk/flix/flix.html} allows the user 
to specify a sky position and obtain upper limits on the EPIC fluxes of a point source 
at the position if the location has been observed by {\em XMM-Newton} but no source was 
detected. The catalogue can also be accessed through
HEASARC\footnote{http://heasarc.gsfc.nasa.gov/db-perl/W3Browse/w3table.pl? tablehead=name\%3Dxmmssc\&Action=More+Options}
and VIZIER\footnote{http://vizier.u-strasbg.fr/cgi-bin/VizieR}.  The results
of the external catalogue cross-correlation carried out for the 3XMM catalogue
(section~\ref{sec:catxcorr}) are available as data products within the XSA and LEDAS or through the XCat-DB.  The {\em XMM-Newton} Survey Science Centre webpages also detail how to provide feedback on the catalogue.

Where the 3XMM catalogue is used for research and publications, please 
acknowledge their use by citing this paper and including the following: \\

{\em This research has made use of data obtained from the 3XMM {\em XMM-Newton} serendipitous source catalogue compiled by the 10 institutes of the {\em XMM-Newton} Survey Science Centre selected by ESA.}

\section{Future catalogue updates}\label{sec:updates}
Incremental releases (data releases) are planned to augment the 3XMM
catalogue.  At least one additional year of data will be included with each data
release. Data release 6 (DR6) will provide data becoming public during 2014 and 2015 and should be released during 2016. These catalogues will be accessible
as described in Section~\ref{sec:access}.

\section{Summary}

This paper presents the third major release of the XMM-Newton serendipitous source 
catalogue (3XMM), in its original version (3XMM-DR4) and in the first incremental 
version (3XMM-DR5). The 3XMM catalogues have been constructed by the {\em XMM-Newton Survey 
Science Centre} and the 3XMM-DR5 catalogue becomes the largest catalogue of X-ray sources 
detected using a single 
X-ray observatory. The characteristics and improvements of this catalogue, with respect 
to previous versions, are outlined as well as how to cite and access the catalogue. This 
paper serves as the reference for future incremental versions of the same catalogue 
(3XMM-DR6, etc), as new {\em XMM-Newton} data becomes publicly available.

\begin{acknowledgements}
Firstly, we extend our thanks to the anonymous referee who raised some very useful points, allowing us to improve this paper. We are also extremely grateful for the strong support provided by the XMM-Newton SOC. We also thank the CDS team for their active contribution and support. The French teams are grateful to Centre National d'\'Etudes Spatiales CNES for their outstanding support for the SSC activities. The German teams are grateful to the Deutsches Zentrum für Luft- und Raumfahrt for supporting this activity under grants 50 OX 1101, 50 OX 1401 and 50 OG 1101. 
The University of Leicester acknowledges the financial support of the UK Space Agency
and also ESA. FJC acknowledges financial support by the Spanish Ministry of Economy and Competitiveness through grant AYA2012-31447, which is partly funded by the FEDER programme, and grant AYA2015-64346-C2-1-P. MTC acknowledges financial
support from the Spanish Ministry of Science and Innovation through grant AYA2010-21490-C02-1.
The Italian team acknowledges financial support during the years from the 
Ministero dell'Istruzione, dell'Universita' e della Ricerca (MIUR), from the 
Agenzia Spaziale Italiana (ASI)  and from the Istituto Nazionale di 
Astrofisica (INAF).

This research has also made use of the SIMBAD database, of the VizieR
catalogue access tool, and of Aladin, operated at CDS, Strasbourg, France, 
the TOPCAT/Stilts software written by Mark Taylor of the University of
Bristol, UK and the NASA HEASARC FTOOLS 
package\footnote{http://heasarc.gsfc.nasa.gov/ftools/} \citep{Blackburn1995}, 
and made extensive use of the ALICE High Performance Computing Facility at the
University of Leicester.

\end{acknowledgements}

\appendix
\section{Known issues affecting 3XMM-DR4 only}\label{ap:knownissues}
{
\begin{itemize}
\item After the creation of the 3XMM-DR4 catalogue, it was discovered that the raw
  event files from the ODFs of a number of mosaic mode sub-pointing
  observations contained corrupted data whereby some of the events in a given
  sub-pointing ODF were actually from another sub-pointing. Since the raw event
  positions are specified in detector coordinates and are subsequently mapped
  to their sky locations during pipeline processing by reference to the
  observation boresight position, which is specified for the given
  sub-pointing, the celestial positions of these events are wrong and
  therefore results in some detections having incorrect celestial
  coordinates. The problem arose in the algorithm used to split the raw parent
  ODF into sub-pointing ODFs.  In some cases all instruments were affected
  while in others, only one or both of the MOS instruments was affected. Of
  the 419 mosaic-mode sub-pointing observations included in 3XMM-DR4, 82 are
  affected to some extent, involving 4918 detections. The affected
  observations are listed in the {\it watchout} section of the XMMSSC 3XMM-DR4
  catalogue web
  pages\footnote{http://xmmssc-www.star.le.ac.uk/Catalogue/xcat\_public\_3XMM-DR4.html}.
  For 3XMM-DR5, none of the affected mosaic sub-pointing observations is 
  included in the catalogue.

\item The vignetting values provided in the 3XMM-DR4 catalogue (for each
  instrument, for bands 1 to 5) were found to have been computed for an energy
  of 0~keV rather than the energy relevant to the band. Thus the values for
  each band of a given instrument are identical. This error does not affect
  the count rates or fluxes as the vignetting correction applied to them is
  computed separately and has been verified as correct. It is only the
  tabulated values in the vignetting columns of the catalogue that are
  incorrect in 3XMM-DR4 and they are correct in 3XMM-DR5.

\item A significant issue identified after the public release of the 3XMM-DR4
  catalogue relates
  to the error values on various quantities.  It was established that the
  error quantities (i.e. columns containing an {\it\_ERR} at the end) for the
  XID band (band 9) count rates and fluxes of a significant number
  ($\sim$42200) of detections ($\sim$10\% of the catalogue) were substantially
  wrong (generally being overestimated by factors up to $\sim$ 100 but in a
  few cases, up to 1000). A more detailed investigation found that while all
  error columns are potentially affected (and therefore also any derived
  parameters involving error-weighted quantities, such as some of the unique
  source quantities), the frequency and magnitude of the problem is much worse
  for the XID band data than any other parameter. It has been established that
  for other key quantities, such as the statistical positional uncertainty
  ({\it RADEC\_ERR}) and the instrument count rates and fluxes in other
  (non-XID) bands, only about 1.3\% of detections are affected and, generally,
  the scale of the problem is very small. For the positional uncertainty,
  1.4\% of detections have incorrect {\it RADEC\_ERR} values and only 0.26\%
  of detections have position errors that differ from their correct values by
  more than 0.05\arcsec\ while for only 89 detections does it differ by more
  than 0.5\arcsec\ (of which, 58 are detected as extended sources and 81 have
  a non-zero quality flag). Furthermore, for 81\% of those detections where
  the position error is wrong by more than $\pm$0.05\arcsec, the correct
  position error is smaller than that quoted in the 3XMM-DR4 catalogue. The most
  extreme deviations of the {\it RADEC\_ERR} values from their correct values are 
  32\arcsec\ larger and 2.3\arcsec\ smaller. For the PN band 2 flux errors, only
  $\sim$1.1\% of detections have values that deviate from their correct
  values by more than $10^{-5}$, when expressed as a fraction of the correct
  value.  For the errors on the XID band photometric quantities (rates,
  fluxes, counts) the correct error is generally smaller than that given in
  3XMM-DR4. 

Thus, while there is a significant problem with the error quantities on the
XID band photometric data in 3XMM-DR4, the problem is much less severe for other
quantities. It is emphasised that the correct error quantities are present in 3XMM-DR5.
\end{itemize}

}

\section{Data modes of XMM-Newton exposures included in the 3XMM catalogue.}\label{ap:datamodes}
{
 The data modes are described in Table~\ref{modestab}.

\begin{table}[t]
\normalsize
\caption{Data modes of XMM-Newton exposures included in the 3XMM catalogue.} 
\label{modestab}
\small
\centering
\begin{tabular}{lll}
\hline \hline
Abbr. & Designation & Description \\
\hline
\multicolumn{3}{l}{\it \ \ MOS cameras:} \\
PFW  &	Prime Full Window & covering full FOV \\
PPW2 &	Prime Partial W2 & small central window \\
PPW3 &	Prime Partial W3 & large central window \\
PPW4 &	Prime Partial W4 & small central window \\
PPW5 &	Prime Partial W5 & large central window \\
FU &	Fast Uncompressed & central CCD in timing mode \\
RFS &   Prime Partial RFS  & central CCD with different frame \\
    &                      & \ \ time (`Refreshed Frame Store') \\
\multicolumn{3}{l}{\it \ \ pn camera:} \\
PFWE  &	Prime Full Window & covering full FOV \\
      &	 \ \ \ Extended & \\
PFW &	Prime Full Window & covering full FOV  \\
PLW &	Prime Large Window & half the height of PFW/PFWE \\[0.2cm]
\hline 
\end{tabular}
\normalsize		
\end{table}
}

\section{Definitions relating to 3XMM-DR5 detections referred to in this work}\label{ap:defs}

We describe here some of the important quantities relating to 3XMM-DR5 detections that are frequently referred to in the paper.

{\it RADEC\_ERR} is the statistical position error, defined as $(ra\_err^{2} + dec\_err^{2})^{1/2}$, where ra\_err and dec\_err are the 1-sigma errors in the RA and DEC coordinate directions, respectively, determined during the fitting of the PSF to the source  image

{\it SYSERRCC} is the estimated 1-sigma error from the rectification process, as defined by equation 3 in section~\ref{sec:rectsyserr}.

{\it POSERR} is the error representing the quadrature combination of {\it RADEC\_ERR} and {\it SYSERRCC}, i.e. $(RADEC\_ERR^{2} + SYSERRCC^{2})^{1/2}  $.

Count rates for detections are given in the <ca>\_<b>\_RATE columns in the catalogue for EPIC camera, <ca>, in energy band <b>, for bands 1-5 and 9. These are the total integrated counts for the detection, derived from within the whole PSF fitted to the source image, divided by the exposure map value at the source position. The count rate values are corrected to the rate on-axis position. The band-8 rates in each camera are the sum over bands 1-5.

Fluxes are provided in the <ca>\_<b>\_FLUX columns. These are converted from the count rates via energy conversion factors (ECFs) (see table 2), assuming a power-law spectrum with $N_H=3\times10^{20}$~cm$^{-2}$ and a power law photon index of 1.7. 

The summary flags, in the {\it SUM\_FLAG} column, provide a simple overview of the quality of the detection, based on a combination of the automatic flags and flags set during manual (visual) screening. Values are: 0 - identifies the best quality detections, i.e. those with no evident complicating factors;  1 - detections where the source parameters may be affected;  2 - cases where the automatic analysis suggests the detection may be a spurious extended source or associated with known detector features but is not flagged by manual screening;  3 - cases where manual screening has flagged the detection but automatic flags are not set;  4 - detections where both automatic and manual screening flags are set.

\section{Detailed description of the issue known as 'missing sources'}\label{ap:miss_srcs}
Of the $\sim$ 25700 missing 3XMM detections, up to 8\% are found only in the pn
band-1 data. Visual inspection of examples and analysis of the pn
detector-image data suggests many of these are probably 
previously unrecognised MIP features, i.e. spurious detections, in 2XMMi-DR3 
(see section~\ref{sec:mips}), though some may well be real, soft
sources. A second, difficult-to-quantify percentage (but $\le$7\%) 
of the missing 3XMM detections may have detected counterparts in the 3XMM
catalogues but be unmatched within 10\arcsec\ due to imperfect astrometry in
either the 2XMMi-DR3 and/or 3XMM catalogue. A third component of up to around
3\% of the missing 3XMM detections may be detections in 2XMMi-DR3 that are
associated with hitherto unrecognised/unflagged detector features - such
features become apparent when the missing 3XMM detections are plotted in
detector coordinates for each EPIC instrument, after allowing for likely real 
detections in the same regions that are detected in more than 1 instrument.

Other explanations for the missing 3XMM detections include 
\begin{itemize}
\item A small number ($<$1\%) are pairs of visually verified close sources 
that were separated in 2XMMi-DR3 but found as either a single extended or a 
single unresolved point source in 3XMM.
\item A small number of cases are likely spurious detections in the wings of
bright sources in 2XMMi-DR3 that were not flagged during the 
manual screening process for 2XMMi-DR3 and were not detected in 3XMM.  
\end{itemize}

The above-mentioned explanations account for $\le$20\% of all the clean,
point-like missing 3XMM detections. Some 75\% of the missing 3XMM detections 
have EPIC likelihoods in 2XMMi-DR3, $L$, $< 10$ (90\% have $L < 15$).  It
might be thought that the missing 3XMM detections could arise from spurious 
detections due to random statistical background fluctuations (false positives) 
in 2XMMi-DR3 - the numbers of such detections, estimated from simulations, was 
discussed in section 9.4 of \cite{wats09}. Using the cumulative count rates 
presented in Fig. 10 of \cite{wats09} and the distribution of exposure times
for observations in 2XMMi-DR3, we estimate  around 7500 detections  in  the
common observations  might  be  false  positives.  This, however, is probably
an overestimate of the contribution of false positives to the missing sources because although there are notable changes to the  pipeline  processing
between the 2XMMi-DR3 and 3XMM catalogues, the input ODFs and associated event 
data are often the same for the common observations, i.e. the data are not 
independent. It  should  be  noted that  of  the $\sim$25700 missing
3XMM-DR5 detections, $\sim$5200 of  them  belong  to  unique sources that
comprise at least one other 2XMM-DR3 detection, hinting that at least 20\% are 
probably real.

The distributions of the band-8 likelihood for clean, point-like EPIC detections found in 3XMM-DR5 and not in 2XMMi-DR3 (and vice versa) are very similar and strongly biased to low ($6<L<10$) likelihood values. Both are much more strongly concentrated in this range than the distribution of all clean, point-like detections. Evidently, the issue of the ’missing’ detections is primarily related to changes affecting detections with likelihoods near the $L=6$ threshold used for the catalogue.

It is clear that two of the major improvements to the 3XMM catalogue with
respect to the 2XMM catalogue, which are the new empirical PSF, described in
Sec.\ref{sec:PSF} and the optimised flare filtering (see
Sec.\ref{sec:optflare}), could have an impact on the detection likelihoods and hence the numbers of detected sources. We investigated the impact of these two
improvements. Optimising the flare filtering maximises the signal to noise ratio of the sources but also affects the background level, as described in Sec.\ref{sec:optflare}. To explore the
impact of changing the background, we scaled the 3XMM-DR5 background  maps
around  their  original  values - raised background  model  values  from
3XMM-DR5  images at the positions of faint sources could reduce the detection
likelihood below the threshold of 6. Our analysis was limited  to  scaling
the  entire  original  background  map  for each available instrument by a
common factor (in steps of 2\% between 90\% and 110\%). While this is not
adequately representative  of real  background variations between processings,
which would vary across the field (see below), it helps to illustrate the 
potential effects that  may  occur.

From  a  subset  of  1854  fields,  we find up to $\sim$9700 extra detections
may appear if the background  is  systematically  underestimated  by  10\%
and $\sim$6800 fewer detections appear if the background is 10\% higher than
the original level. However, as noted, the differences in the background maps
between the 2XMMi-DR3 and 3XMM processings are much more likely to occur on a
spatially localised scale rather than a uniform change  across  the  field  of
view. To  look  for indicators that this might be the case, 
we computed ratios of the background  maps  (3XMM-DR5/2XMMi-DR3)  in  each  
instrument and
band and looked for deviations of the ratio (relative to the mean of the ratio
image) at the positions of  2XMMi-DR3  detections that  are  missing  in
3XMM-DR5. We observe a spread of up to $\pm$20\% in the deviations of the
ratio at some source positions but no evidence of systematic background
over-estimation in a specific instrument or band. Nevertheless, background 
enhancements that push the EPIC band 8 detection likelihood below 6 could be
arising in a different instrument and/or energy band in each case.

\begin{figure}
   \centering
   \includegraphics[width=8cm, angle=-90]{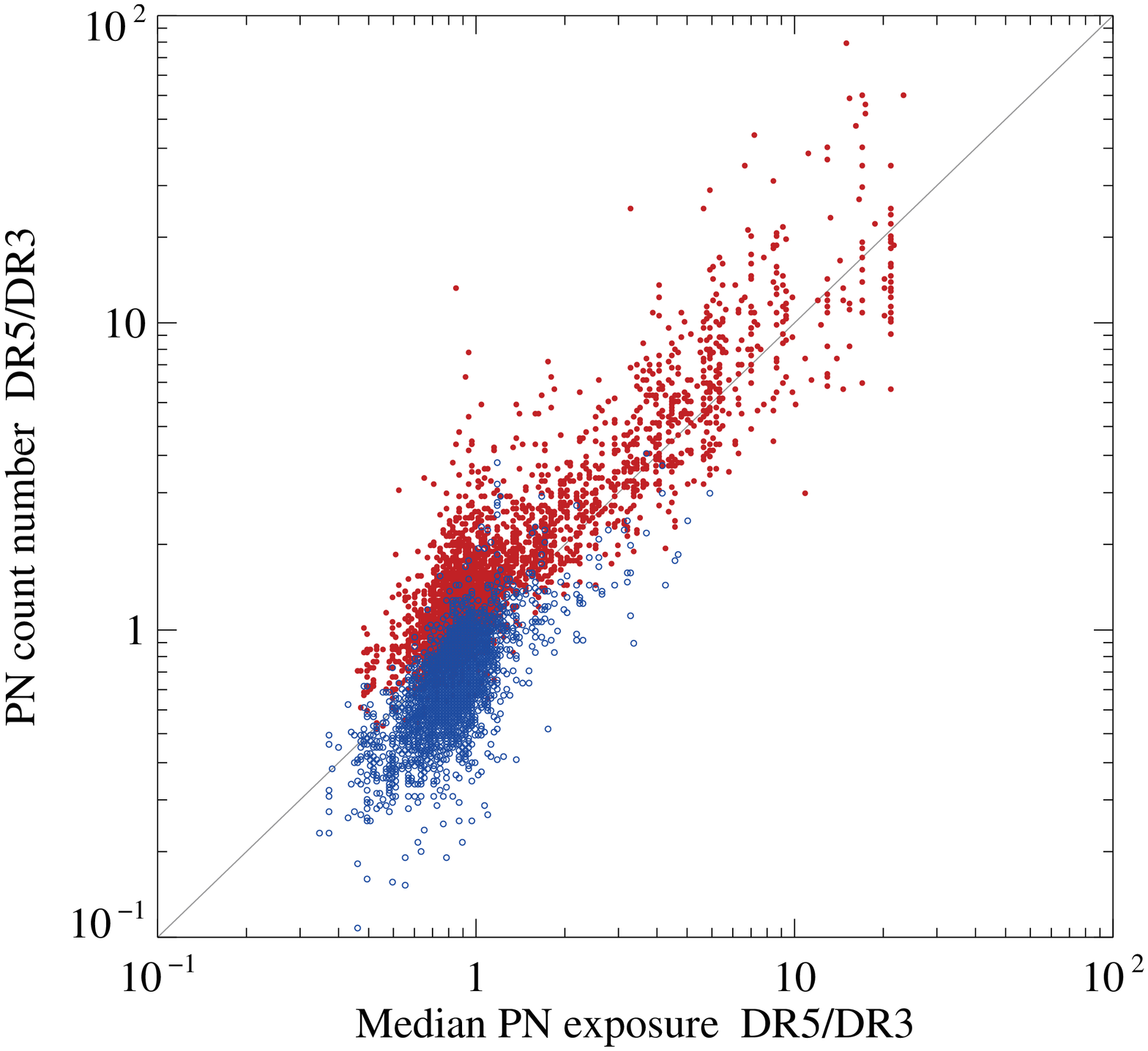}
      \caption{Ratio of the pn source count numbers (DR5/DR3) plotted against the corresponding ratio of median pn exposure times. Red filled points are pn detection likelihood in 3XMM-DR5, $L_{pn(DR5)}$ $>6$ and pn detection likelihood in 2XMMi-DR3, $L_{pn(DR3)}$ $< 6$. The blue open points are $L_{pn(DR5)} < 6$ and $L_{pn(DR3)} > 6$.}
         \label{fig:counts_exposure}
   \end{figure}

The second effect of the improved flare filtering is the impact on the good time intervals. We investigated the relation between exposure time, the number of counts in the source (count number) and detection likelihood, using only the pn parameters of detections that are in both 2XMMi-DR3 and 3XMM-DR5 (i.e. whose EPIC detection likelihood is $>6$ in each catalogue). We expect that we would see a similar relation for the combined instrument (EPIC) source parameters if we could include detections with EPIC likelihoods below six in the catalogues (which, by definition,  we don't have). Fig.~\ref{fig:counts_exposure} show the ratio of pn source count numbers (DR5/DR3) plotted against the corresponding ratio of median pn exposure times. The data include only point sources with {\it SUM\_FLAG}~$\le 1$ that are isolated and not affected by nearby extended sources, in both catalogues. The red points reflect sources whose pn detection likelihood in 3XMM-DR5, $L_{pn(DR5)}$, is $>6$ while their  pn detection likelihood in 2XMMi-DR3, $L_{pn(DR3)}$, is $< 6$ - these are detections that, based on their pn data, would be present in 3XMM-DR5 and not in 2XMMi-DR3. The blue points represent data where $L_{pn(DR5)} < 6$ and $L_{pn(DR3)} > 6$, i.e. which would be in 2XMMi-DR3 and not in 3XMM-DR5.

\begin{figure}
   \centering
   \includegraphics[width=9cm]{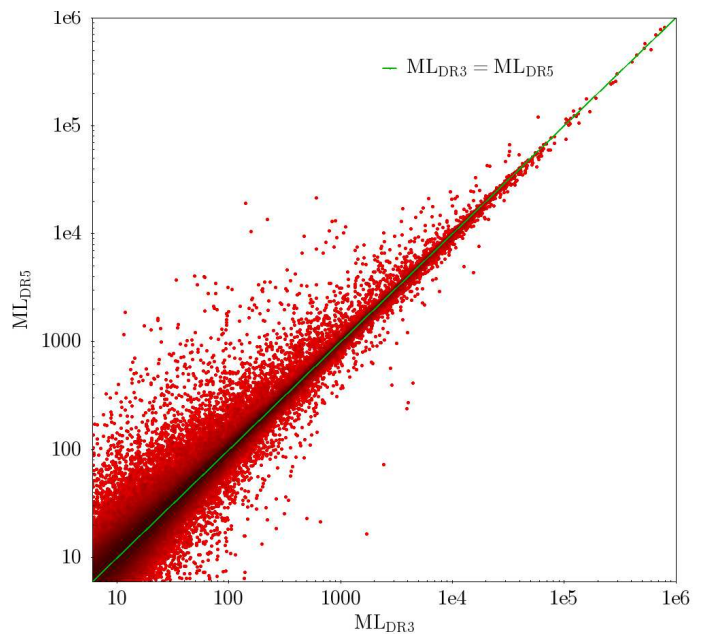}
      \caption{The maximum likelihood (ML) values of all the sources common to 3XMM-DR5 and 2XMMi-DR3. On the ordinate are the ML values in 3XMM-DR5 and on the abscissa, the ML values in 2XMMi-DR3. The (green) solid diagonal line indicates  where the 3XMM-DR5 and 2XMMi-DR3 ML values are equal.}
         \label{fig:improveML}
   \end{figure}

To investigate the impact of changing the PSF to the empirical model, we
reprocessed the 4921 fields that are common to 3XMM-DR5 and 2XMMi-DR3 using
the same source-detection steps of the pipeline, input data and calibration 
files that were employed to create 3XMM-DR5, but changing the new
empirical PSF model to the previous 'medium-accuracy' model. We found some 
8300 clean, point-like 3XMM-DR5 detections have no matching detection obtained
with the medium-accuracy PSF within 5\arcsec\ (7100 within 10\arcsec), 
demonstrating that the new PSF has a non-negligible
effect on the source detection. It should also be pointed out that {\em
  emldetect} gives the most reliable results when the PSF model used is
similar to the true PSF \citep{feig06}, implying that the empirical PSF, constructed from observed source data, should provide more reliable sources.  Changing both the background and PSF therefore
has an impact on the maximum likelihood determined. Fig.~\ref{fig:improveML}
shows the relationship between the maximum likelihood (ML) in  3XMM-DR5 and
2XMMi-DR3 for all sources common to both catalogues. More than half the
sources have a higher maximum likelihood in 3XMM-DR5 compared to 2XMMi-DR3,
indicating generally better sensitivity in 3XMM-DR5. It should also be noted that, as indicated in the {\em emldetect} description\footnote{http://xmm.esac.esa.int/sas/current/doc/emldetect/node3.html}, the maximum likelihood values provide only a rough estimate of the number of expected spurious sources for low count sources \citep[$\lesssim$ 9,][]{cash79}. As many as 10\% of the catalogue sources have counts $<$ 9 in at least one instrument, so sources with a low count rate may have an inaccurate likelihood value attributed.

The changes to the pipeline generating the catalogue sources mean that the maximum likelihood value attributed to each source varies from catalogue to catalogue. The dispersion is high ($\sim$2 in ML) for the distribution of ML values in one catalogue, given a specific ML value in the other catalogue. Given that we have chosen a threshold of ML $\ge$ 6 to indicate a real source, many sources with a ML close to this value in one catalogue will have an ML $<$ 6 in the other, due to this broad dispersion. Indeed, as many as 10000 sources can be found below this threshold in the other catalogue (2XMMi-DR3 when comparing with 3XMM-DR5 or 3XMM-DR5 when comparing with 2XMMi-DR3), and are therefore considered as missing when considering one catalogue over the other. In reality, it is simply that the ML value has fallen slightly below our chosen threshold and therefore the source is just not included in the catalogue.   

In conclusion, the main reason for the missing sources is that the changes in the pipeline processing procedure between 2XMMi-DR3 and 3XMM-DR5 have produced slightly different likelihood values (of mostly real sources), so that detections near the likelihood threshold cross the boundary, in both directions, between catalogues, resulting in different lists of detections. Overall, however, the 3XMM-DR5 procedure is better, resulting in more sources.

\section{Astrometry and the deviation from the Rayleigh distribution}\label{ap:ast_rayleigh}

  To explore the cause(s) of the deviations from the Rayleigh curve, we
  first examined whether the outlier pairs in the tail excess might be
  spurious XMM-quasar associations, though as noted by \cite{wats09}, the
  false match rate for quasars is expected to yield far fewer mismatches than
  the numbers found in the tail excess. To test this possibility we compared
  the distribution of the 3XMM-DR5 EPIC band 8 flux ($F_X$) to SDSS (r band)
  flux ($F_r$) ratio (i.e. $F_X / F_r$) for XMM-quasar pairs from $x > 3.5$
  (the region of the excess tail where the Rayleigh function predicts
  negligible numbers) to that from $x < 0.8$ (where the data and model match
  well). While pairs from the tail do have a slightly (25\%) higher
  $F_{X}/F_{r}$ ratio on average than those at $x < 0.8$ it is too small a
  difference to be explained by systematic mismatching. This conclusion is
  supported by considering XMM-quasar pairs in the tail whose X-ray detections
  belong to 3XMM-DR5 unique sources that include one or more other X-ray
  detections with a quasar counterpart.  Amongst 104 such unique sources
  involving an XMM-quasar pair from the tail, only 13 are cases where all other
  constituent XMM-quasar pairs have $x> 3.5$. These 13 cases might reflect
  mismatches of the XMM detections and quasars. However, the X-ray detections
  involved represent only 8\% of the X-ray detections with quasar counterparts
  that make up the 104 unique sources, suggesting most of the XMM-quasar pairs
  from the tail are not mismatches.

We then constructed distributions for many XMM catalogue parameters (e.g. 
position errors, off-axis angle, count rates, equatorial and galactic 
location, exposure times, nearest-neighbour distance etc.), comparing the
distributions of the data subsets from $x>3.5$ and $x<0.8$. The 
position error ({\it POSERR}) distributions of the two subsets are very
similar while the XMM-quasar separations are markedly different, having 
an average of 0.4\arcsec\ for data with $x<0.8$ compared to 5.5\arcsec\ 
for the $x>3.5$ subset. There is an indication that the points at $x > 3.5$ 
tend to lie at larger off-axis angles. No other trends could be discerned 
from the distributions for other parameters. To push this further, we also 
cross-matched the 6614
3XMM-DR5 detections with SDSS quasar counterparts against the Chandra
catalogue \citep{evans2014}. Within 10\arcsec, 745 XMM detections have one or
more matches with Chandra detections - we retained only the nearest match
in the few instances where multiple matches were present. 
The 3XMM-DR5 detections from the tail do tend to be notably 
more offset from their Chandra counterparts compared to those detections with 
$x < 3.5$. Furthermore, although numbers are more limited, for the XMM-quasar 
pairs in the tail with Chandra counterparts, the error-normalised offsets 
between the Chandra detections and the SDSS quasars appear to provide a better
match to the Rayleigh distribution - there is no evidence of a similar tail 
excess. This hints at the positions of the 3XMM-DR5 detections in the tail 
being incorrect. While, alternatively, their position errors may be 
underestimated, if so, there is no clear evidence the errors are being 
systematically underestimated, e.g. being incorrectly characterised with 
off-axis angle. It is worth noting that while the proportion is lower in the 
central regions of the field of view, even in the $10' < \theta < 12'$
annulus ($\theta$ is the EPIC off-axis angle), 5.6\% of the XMM-quasar 
pairs have $x > 3.5$ - this demonstrates there is not a generic problem with sources at higher off-axis angles.

As noted, there is an indication that XMM sources at $x > 3.5$ tend to lie 
at higher off-axis angles in the field than those from lower $x$ values. 
This is illustrated in figure~\ref{fig:rayleigh_corr} where, alongside the 
histogram of error-normalised offsets for all the XMM-quasar pairs (red), 
we show the histograms for data from off-axis angles $\theta <
5'$ (blue), $5'\le\theta<10'$ (green) and $10'\le\theta<15'$ (grey). For 
sources near the centre of the field, the distribution peaks
too early, at $x=\Delta{r}/\sigma_{tot}\sim0.65$ but better matches the tail 
at $x > 2.5$. Conversely, data from $10' \le \theta < 15'$ peak near $x=1$ but 
account for much of the excess tail. We examined whether this could arise from,
for example, an error in the rotation correction of the rectification
process (see section~\ref{sec:rectfrcorr}) in some observations. If so, for a given field, one might anticipate 
the quasar counterparts having a systematic offset, either ahead of, or 
behind, the X-ray detection, in a sector oriented perpendicular to the radial 
vector, {\bf r}, from the field boresight to the X-ray position. This is not, 
however, evident in fields that contain useful numbers (up to 22) of
XMM-quasar pairs, one or more of which come from $x >3.5$. We also performed 
a more detailed analysis in which the circularised statistical XMM position 
error, {\it RADEC\_ERR}, used previously was
replaced with an error derived from an error ellipse: an elliptical error
contour should better characterise positional uncertainties arising from the 
elongated PSF profiles that become evident at larger off-axis angles. 
Assuming the ellipse is oriented with the major axis tangential to {\bf r}, the mean geometry of the error ellipse as a function of off-axis angle was obtained via the separate errors in RA and DEC of all 3XMM-DR5 detections (available in the initial emldetect source lists, though only the circularised RADEC\_ERR value is provided in the final observation summary source lists). For each XMM-quasar pair, the idealised error ellipse of the X-ray source was scaled to the measured RA and DEC errors and a mean of the major and minor axes was obtained. Using the mean elliptical 
positional uncertainty to normalise the XMM-quasar offsets, even when combined 
with the elliptical errors for the rectification correction and the elliptical 
quasar position errors, still results in a notable excess at $x > 3.5$.

We conclude that the excess of 3XMM-DR5 detections with error-normalised
offsets from their SDSS quasar counterparts $> 3.5$ appears to have a modest 
dependence on the off-axis location of the detection in the XMM field of view, with
a small fraction of detections at higher off-axis angles having either 
incorrect positions or underestimated errors, while sources near the centre 
may have slightly overestimated errors. However, no systematic cause is identified.


\bibliographystyle{bibtex/aa} 

\bibliography{3XMM_v16}




\end{document}